\documentclass[twocolumn]{aastex631}

\shorttitle{OGLE-2014-BLG-0221}
\shortauthors{Kirikawa et al.}
\usepackage{comment}
\usepackage{here}
\graphicspath{{./}{plot/}}
\usepackage{afterpage}
\usepackage{subfigure}
\usepackage{bm}

\begin{document}

\title{\large OGLE-2014-BLG-0221Lb: A Jupiter Mass Ratio Companion Orbiting either a Late-Type Star or a Stellar Remnant}

\author{Rintaro Kirikawa}
\affiliation{Department of Earth and Space Science, Graduate School of Science, Osaka University, Toyonaka, Osaka 560-0043, Japan}
\affiliation{MOA collaboration}
\author{Takahiro Sumi}
\affiliation{Department of Earth and Space Science, Graduate School of Science, Osaka University, Toyonaka, Osaka 560-0043, Japan}
\affiliation{MOA collaboration}
\author{David P.~Bennett}
\affiliation{Code 667, NASA Goddard Space Flight Center, Greenbelt, MD 20771, USA}
\affiliation{Department of Astronomy, University of Maryland, College Park, MD 20742, USA}
\affiliation{MOA collaboration}
\author{Daisuke Suzuki}
\affiliation{Department of Earth and Space Science, Graduate School of Science, Osaka University, Toyonaka, Osaka 560-0043, Japan}
\affiliation{MOA collaboration}
\author{Naoki Koshimoto}
\affiliation{Department of Earth and Space Science, Graduate School of Science, Osaka University, Toyonaka, Osaka 560-0043, Japan}
\affiliation{MOA collaboration}
\author{Shota Miyazaki}
\affiliation{Institute of Space and Astronautical Science, Japan Aerospace Exploration Agency, 3-1-1 Yoshinodai, Chuo, Sagamihara, Kanagawa 252-5210, Japan}
\affiliation{MOA collaboration}
\author{Ian A. Bond}
\affiliation{Institute of Natural and Mathematical Sciences, Massey University, Auckland 0745, New Zealand}
\affiliation{MOA collaboration}
\author{Andrzej Udalski}
\affiliation{Warsaw University Observatory, Al.~Ujazdowskie~4, 00-478~Warszawa, Poland}
\affiliation{OGLE collaboration}
\author{Nicholas J. Rattenbury}
\affiliation{Department of Physics, University of Auckland, Private Bag 92019, Auckland, New Zealand}
\affiliation{MOA collaboration}

\collaboration{20}{(Leading Authors)}

\author{Fumio Abe}
\affiliation{Institute for Space-Earth Environmental Research, Nagoya University, Nagoya 464-8601, Japan}
\author{Richard Barry}
\affiliation{Code 667, NASA Goddard Space Flight Center, Greenbelt, MD 20771, USA}
\author{Aparna Bhattacharya}
\affiliation{Code 667, NASA Goddard Space Flight Center, Greenbelt, MD 20771, USA}
\affiliation{Department of Astronomy, University of Maryland, College Park, MD 20742, USA}
\author{Hirosane Fujii}
\affiliation{Department of Earth and Space Science, Graduate School of Science, Osaka University, Toyonaka, Osaka 560-0043, Japan}
\author{Akihiko Fukui}
\affiliation{Department of Earth and Planetary Science, Graduate School of Science, The University of Tokyo, 7-3-1 Hongo, Bunkyo-ku, Tokyo 113-0033, Japan}
\affiliation{Instituto de Astrof\'isica de Canarias, V\'ia L\'actea s/n, E-38205 La Laguna, Tenerife, Spain}
\author{Ryusei Hamada}
\affiliation{Department of Earth and Space Science, Graduate School of Science, Osaka University, Toyonaka, Osaka 560-0043, Japan}
\author{Yuki Hirao}
\affiliation{Institute of Astronomy, Graduate School of Science, The University of Tokyo, 2-21-1 Osawa, Mitaka, Tokyo 181-0015, Japan}
\author{Stela Ishitani Silva}
\affiliation{Department of Physics, The Catholic University of America, Washington, DC 20064, USA}
\affiliation{Code 667, NASA Goddard Space Flight Center, Greenbelt, MD 20771, USA}
\author{Yoshitaka Itow}
\affiliation{Institute for Space-Earth Environmental Research, Nagoya University, Nagoya 464-8601, Japan}
\author{Yutaka Matsubara}
\affiliation{Institute for Space-Earth Environmental Research, Nagoya University, Nagoya 464-8601, Japan}
\author{Yasushi Muraki}
\affiliation{Institute for Space-Earth Environmental Research, Nagoya University, Nagoya 464-8601, Japan}
\author{Greg Olmschenk}
\affiliation{Code 667, NASA Goddard Space Flight Center, Greenbelt, MD 20771, USA}
\author{Cl\'ement Ranc}
\affiliation{Sorbonne Universit\'e, CNRS, UMR 7095, Institut d'Astrophysique de Paris, 98 bis bd Arago, 75014 Paris, France}
\author{Yuki K. Satoh}
\affiliation{Department of Earth and Space Science, Graduate School of Science, Osaka University, Toyonaka, Osaka 560-0043, Japan}
\affiliation{College of Science and Engineering, Kanto Gakuin University, 1-50-1 Mutsuurahigashi, Kanazawa-ku, Yokohama, Kanagawa 236-8501, Japan}
\author{Mio Tomoyoshi}
\affiliation{Department of Earth and Space Science, Graduate School of Science, Osaka University, Toyonaka, Osaka 560-0043, Japan}
\author{Paul . J. Tristram}
\affiliation{University of Canterbury Mt.\ John Observatory, P.O. Box 56, Lake Tekapo 8770, New Zealand}
\author{Aikaterini Vandorou}
\affiliation{Code 667, NASA Goddard Space Flight Center, Greenbelt, MD 20771, USA}
\affiliation{Department of Astronomy, University of Maryland, College Park, MD 20742, USA}
\author{Hibiki Yama}
\affiliation{Department of Earth and Space Science, Graduate School of Science, Osaka University, Toyonaka, Osaka 560-0043, Japan}
\author{Kansuke Yamashita}
\affiliation{Department of Earth and Space Science, Graduate School of Science, Osaka University, Toyonaka, Osaka 560-0043, Japan}

\collaboration{40}{(MOA \scriptsize COLLABORATION)}

\author{Przemek Mr{\'o}z}
\affiliation{Warsaw University Observatory, Al.~Ujazdowskie~4, 00-478~Warszawa, Poland}
\author{Rados{\l}aw Poleski}
\affiliation{Warsaw University Observatory, Al.~Ujazdowskie~4, 00-478~Warszawa, Poland}
\author{Jan Skowron}
\affiliation{Warsaw University Observatory, Al.~Ujazdowskie~4, 00-478~Warszawa, Poland}
\author{Micha{\l} K. Szyma{\'n}ski}
\affiliation{Warsaw University Observatory, Al.~Ujazdowskie~4, 00-478~Warszawa, Poland}
\author{Igor Soszy{\'n}ski}
\affiliation{Warsaw University Observatory, Al.~Ujazdowskie~4, 00-478~Warszawa, Poland}
\author{Pawe{\l} Pietrukowicz}
\affiliation{Warsaw University Observatory, Al.~Ujazdowskie~4, 00-478~Warszawa, Poland}
\author{Szymon Koz{\l}owski}
\affiliation{Warsaw University Observatory, Al.~Ujazdowskie~4, 00-478~Warszawa, Poland}
\author{Krzysztof Ulaczyk}
\affiliation{Warsaw University Observatory, Al.~Ujazdowskie~4, 00-478~Warszawa, Poland}
\affiliation{Department of Physics, University of Warwick, Gibbet Hill Road, Coventry, CV4~7AL,~UK}

\collaboration{8}{(OGLE \scriptsize COLLABORATION)}



\begin{abstract}
We present the analysis of microlensing event OGLE-2014-BLG-0221, a planetary candidate event discovered in 2014. The photometric light curve is best described by a binary-lens single-source model. Our light curve modeling finds two degenerate models, with event timescales of $t_\mathrm{E}\sim70$ days and $\sim110$ days. These timescales are relatively long, indicating that the discovered system would possess a substantial mass. The two models are similar in their planetary parameters with a Jupiter mass ratio of $q \sim 10^{-3}$ and a separation of $s \sim 1.1$. While the shorter timescale model shows marginal detection of a microlensing parallax signal, the longer timescale model requires a higher order effect of microlensing parallax, lens orbital motion or xallarap to explain the deviation in the light curve. However, the modeling shows significant correlation between the higher order effects and suffers the ecliptic degeneracy that results in a failure to determine the parallax parameters. Bayesian inference is used to estimate the physical parameters of the lens, revealing the lens to be either a late-type star supported by the shorter timescale model or a stellar remnant supported by the longer timescale model. If the lens is a remnant, this would be the second planet found by microlensing around a stellar remnant. Since the models predict different values for relative proper motion and source flux, future high angular resolution follow-up observations  (e.g. Keck adaptive optics) are required to rule out either of the models.
\end{abstract}



\section{Introduction} \label{sec:intro}
Gravitational microlensing is a well-known method for the discovery of exoplanets. Since the first discovery of a planet via microlensing in 2004 \citep{Bond+2004,Bennett+2006}, the number of exoplanets detected by microlensing has grown to 200 \citep{Akeson+2013}. 
The most likely lens star in a microlensing event is a low mass late type star as these are the most prolific in our Galaxy. A large fraction of planets discovered by microlensing are gas giants\footnote{110 discovered microlensing planets have estimated mass larger than Saturn \citep{Akeson+2013}. Note that microlensing has higher detection efficiency for higher mass ratio planets.}. Such planets are less likely to be formed around low mass stars according to the core accretion model of planet formation \citep{Ida+2005, Burn+2021}. 
\cite{Suzuki+2016} gave a statistical analysis of the set of microlensing planets and found the planet-star mass ratio function was best characterized by a broken power-law with the break at $q=1.7\times10^{-4}$ which corresponds to the mass of Neptune. Comparison of the result to population synthesis models of the core accretion admitted excess in the microlensing planet beyond $q \sim 10^{-4}$ \citep{Suzuki+2018}, including a gas giant mass ratio regime. \par
Microlensing surveys were originally initiated to search for dark matter in the form of MAssive Compact Halo Objects \citep[MACHOs;][]{Paczynski+1986} in which stellar remnants, substellar objects and planets are the expected populations. Identifying isolated black holes as a part of a population of MACHOs was not successful until recently owing to the degeneracy in mass and distance present in the interpretation of most single lens events. The first definitive detection of an isolated black hole was made possible through observations of the astrometric shift of the apparent source position using high resolution imaging of the Hubble Space Telescope \citep{Sahu+2022, Lam+2022, Mroz+2022}. \par
The first detection of a stellar remnant by microlensing was made in 2021 \citep{Blackman+2021}. Following the actual discovery in 2010 \citep{Bachelet+2012}, follow-up observations with high resolution imaging using the Keck-I\hspace{-.1em}I telescope were conducted to resolve the source and lens and resulted in no detection of a luminous lens. The lens system was concluded to be composed of a white dwarf and accompanying Jovian planet. \par
So far, these are the only examples of stellar remnants found by microlensing while there are a few more candidates to be confirmed \citep[e.g.][]{Miyake+2012, Shvarzvald+2015}. An ambitious program with the Nancy Grace Roman Space Telescope, the NASA's next flagship mission \citep{Spergel+2015}, has been proposed to detect isolated black holes, utilizing Roman's high precision photometry and astrometry \citep{Lam+2023}. The survey strategy required for detection of isolated black holes is mostly satisfied by the notional design of the Galactic Bulge Time Domain Survey optimized for detection of exoplanets toward the Galactic bulge. \cite{Lam+2023} predicts more than 300 isolated black hole detections among which 270 black holes can be characterized by additional daily cadence observations and an astrometric precision of 0.1 mas. The same strategy should also enable the discovery of black hole binaries. These are less common but more likely to be characterized with additional constraints on the mass distance degeneracy from accurate measurement of a projected source size. A future black hole survey using the powerful capability of Roman seems promising; however, we need to wait until its expected launch in 2026. A discussion now of the discoveries of black holes and other compact objects is important for establishing the future prospects of such a survey.  \par
This paper presents analysis of microlensing event OGLE-2014-BLG-0221. Section \ref{sec:observation} describes how the event was discovered and observed. Section \ref{sec:modelling} gives the details of our light curve analysis. In section \ref{sec:CMD}, we investigate the source property using a color magnitude diagram for field stars around the event coordinate. The lens properties are estimated in section \ref{sec:Bayesian}. Then, we discuss the results in section \ref{sec:discussion}.

\section{Observation} \label{sec:observation}
Microlensing event OGLE-2014-BLG-0221 was detected by the Optical Gravitational Lensing Experiment 
\citep[OGLE;][]{Udalski+2015} 
Collaboration on 2014 March 6 ($\mathrm{HJD'}\equiv\mathrm{HJD}-2450000\sim6723$) as a part of the OGLE-IV survey. 
The OGLE group conducted the observations with their 1.3m Warsaw telescope at the Las Campanas Observatory in Chile. 
The event coordinate was first reported at $\rm{(RA,Dec)_{J2000}}=(18h01m12.90s, -27^\circ25'37.2'')$, corresponding to Galactic coordinates $(l,b)=(3.044660, -2.195338)$, where a known faint star is located; however later, a centroid shift of the lensed source was observed while the source was magnified. 
The coordinate was corrected to $\rm{(RA,Dec)_{J2000}}=(18h01m12.90s, -27^\circ25'36.35'')$, corresponding to Galactic coordinates $(l,b)=(3.044658,-2.195316)$, and the event was re-identified by the New Objects in the OGLE Sky (NOOS) system \citep{Udalski+2003} as the lensing of a previously unknown object and re-designated as OGLE-2014-BLG-0284.
OGLE-2014-BLG-0221 was located in the OGLE field BLG511 and was detected using a nominal cadence of every 60 minutes for their \textit{I}-band observations. \par

The event was also independently discovered by the Microlensing Observations in Astrophysics 
\citep[MOA;][]{Bond+2001, Sumi+2003} 
Collaboration as a part of the MOA-I\hspace{-.1em}I survey on 2014 March 9 and designated as MOA-2014-BLG-069. 
The event was located in MOA field gb10. This field was one which was observed in the survey with a cadence of 15 minutes. 
The MOA group uses their 1.8m MOA-I\hspace{-.1em}I telescope and $2.2\ \mathrm{deg}^2$ wide field of view camera, MOA-cam3, at the Mount John Observatory in New Zealand \citep{Sako+2008}. \par

The OGLE photometric data were obtained mostly in the standard Kron-Cousins \textit{I}-band and occasionally in the standard Johnson \textit{V}-band in order to extract color information of the source star. 
The MOA photometric data were obtained in the designated MOA-Red band, equivalent to a combined band of the standard Cousins \textit{R} and \textit{I}. MOA also occasionally observes in Johnson\textit{V}-band; however unfortunately, no data were taken in 2014.
OGLE and MOA reduced the data with their own pipelines \citep{Udalski+2003, Bond+2001} based on their implementation of the difference image analysis method \citep{Alard+1998}. \par

The pipelines often underestimate uncertainties of photometric data for a stellar dense region. 
For this reason, we rescaled the errors to account for low level unknown systematics so that the reduced $\chi^2$ (or $\chi^2/d.o.f.$) equals 1, following the standard procedure detailed in \citep{Bennett+2008, Yee+2012}. The rescaling formula is 
\begin{eqnarray}
\sigma'_{i} = k\sqrt{\sigma^{2}_{i}+e^{2}_\mathrm{min}},
\end{eqnarray}
where $\sigma'_{i}$ is the rescaled error, $\sigma_{i}$ is the error before rescaling, and $k$ and $e^{2}_\mathrm{min}$ are the rescaling parameters. Data sets and rescaling parameters adopted in the analysis are listed in Table \ref{tab:data}. We used the optimized photometry of OGLE-2014-BLG-0221 for OGLE data.

\begin{table}[hbtp]
  \caption{Data Set and Rescaling Parameters}
  \label{tab:data}
  \centering
  \begin{tabular}{cccccc}
    \hline \hline
    Name & Telscope & Filter & $N_\mathrm{data}\footnote{The number of data points.}$ & $k$\footnote{The rescaling parameters.\label{foot:1}} & $e_\mathrm{min}\footref{foot:1}$ \\
    \hline 
    OGLE & Warsaw 1.3 m & \textit{I} & $2480$ & $1.575$ & $0$ \\
     & & \textit{V} & $205$ & $1.336$ & $0.003$ \\
    MOA & MOA-I\hspace{-.1em}I 1.8 m & MOA-Red & $7835$ & $0.974$ & $0.026$ \\
    \hline
  \end{tabular}
\end{table}

\begin{figure}[ht!]
\includegraphics[width=\columnwidth]{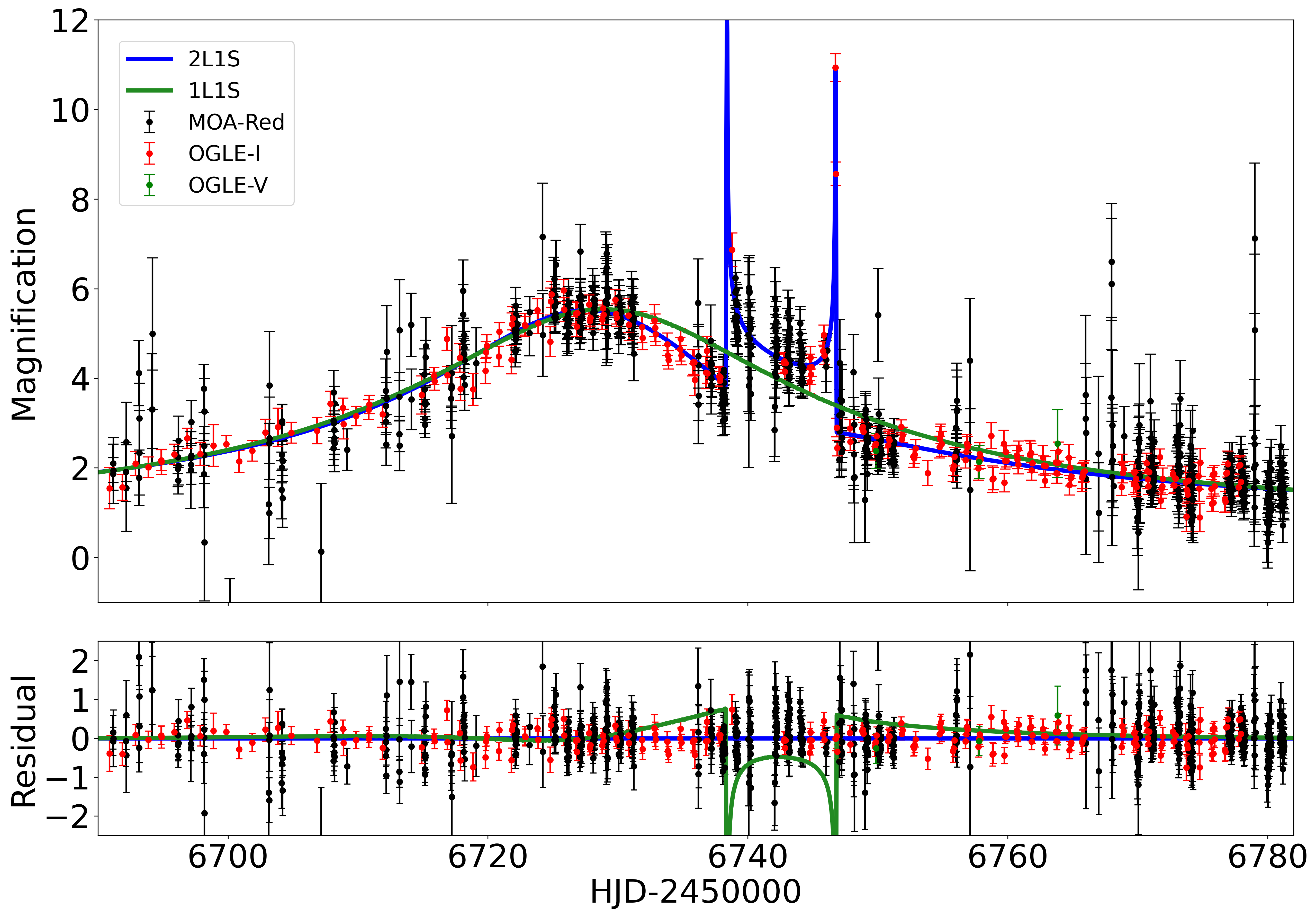}
\caption{The light curve of OGLE-2014-BLG-0221. The black, red and green data points are the reduced photometric data from the MOA-Red, OGLE \textit{I}- and \textit{V}-band observations, respectively. The blue and green lines represent the best-fit binary lens single source model and best-fit single lens single source model. The lower panel shows the residuals from the best-fit binary lens model.\label{fig:1}}
\end{figure}

\begin{figure*}[ht!]
\includegraphics[width=\linewidth]{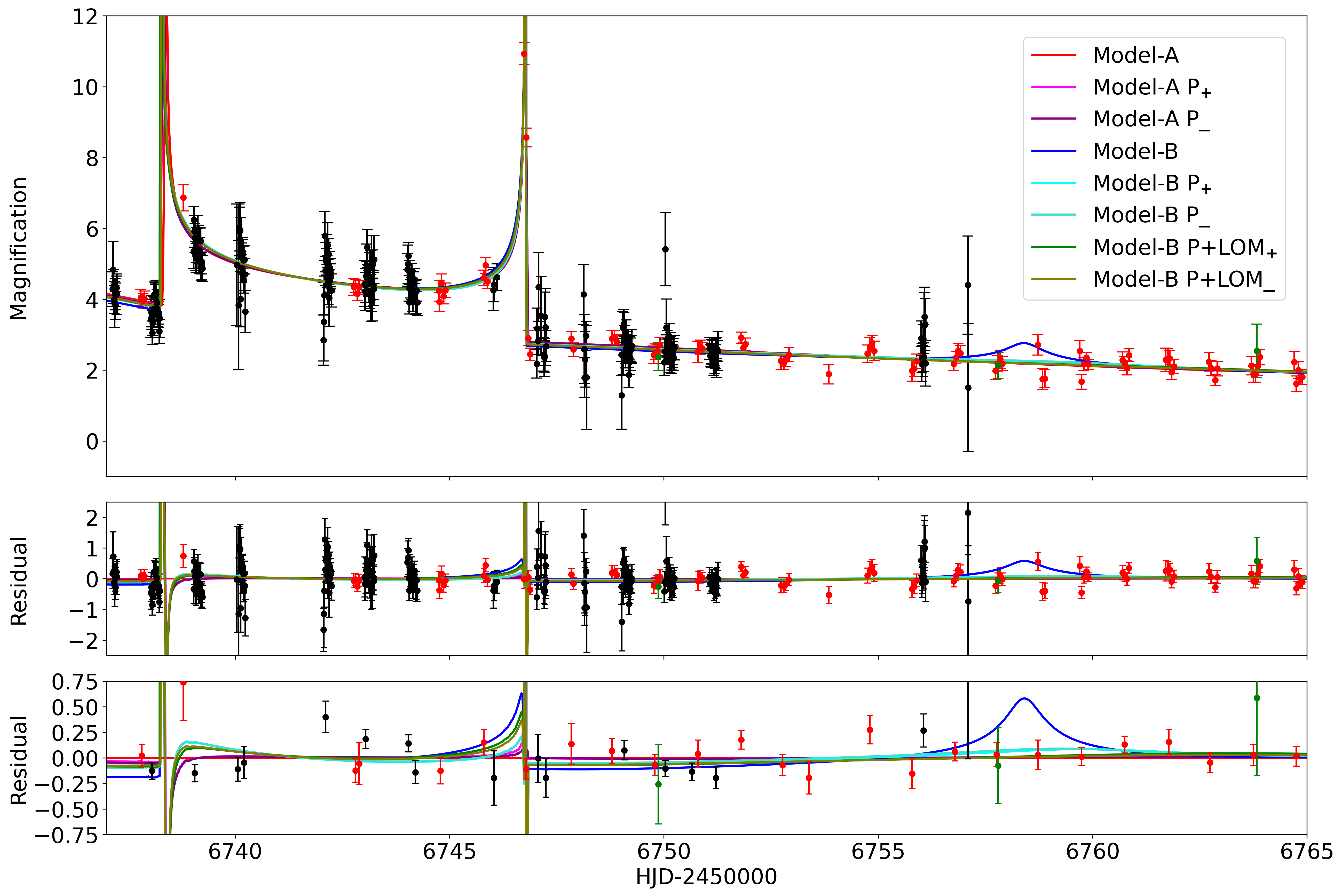}
\caption{Model light curves of the two degenerate models around the anomaly. The Model-A, Model-A P$_\pm$, Model-B, Model-B P$_\pm$ and Model-B P+LOM$_\pm$ are plotted. Since different source and blending fluxes between the Model-A and Model-B make different magnification, magnification of the Model-B is scaled by source and blending fluxes of the Model-A for comparison. The lower residual panel shows the data points binned in 1 day interval.\label{fig:2}}
\end{figure*}
%
%
\begin{figure*}
  \hspace{-30pt}
  \begin{minipage}{0.35\textwidth}
    \centering
        \subfigure[Model-A]{\includegraphics[width=8cm]{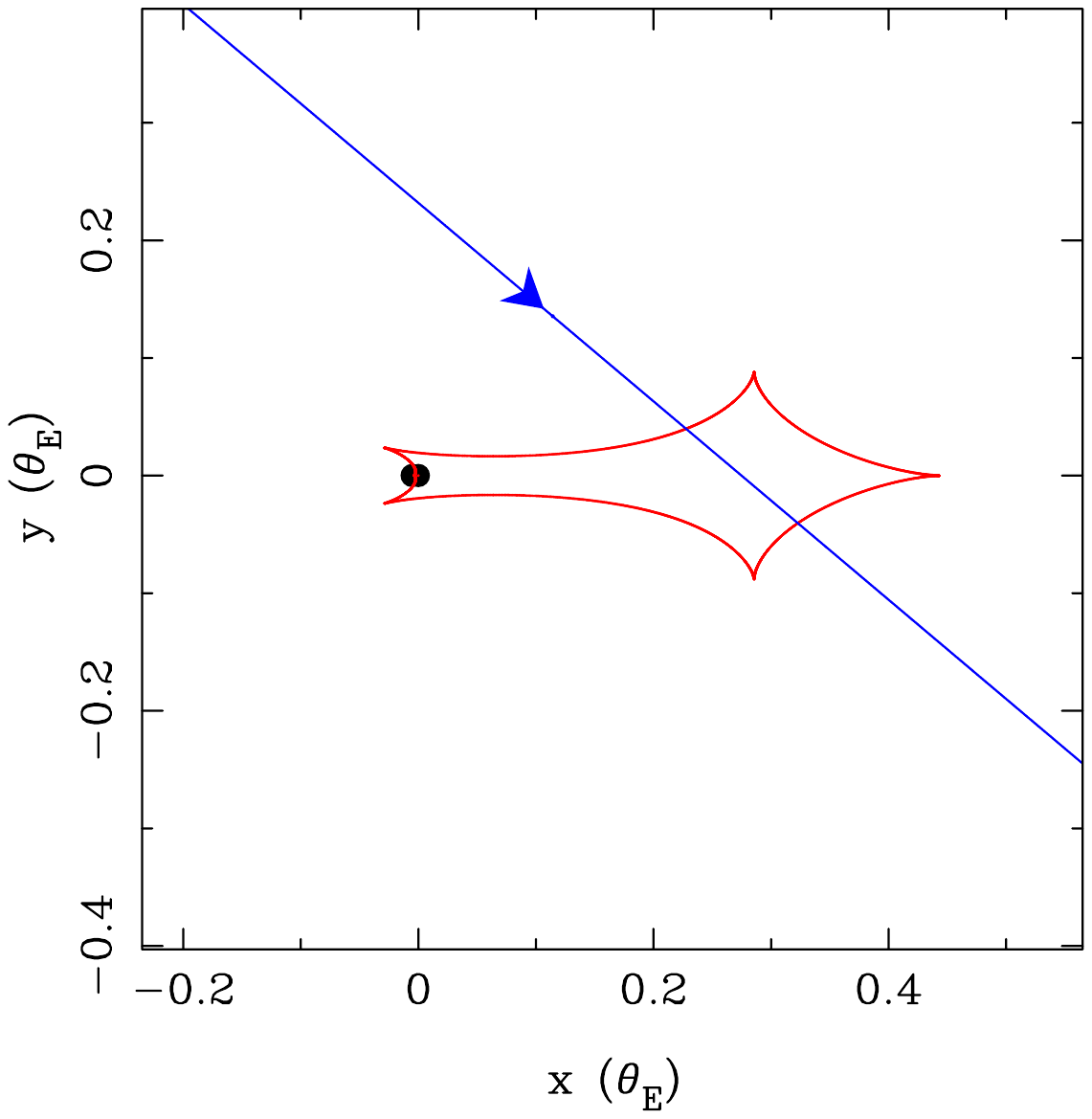}}
  \end{minipage}\hspace{-20pt}
  \begin{minipage}{0.35\textwidth}
    \centering
        \subfigure[Model-A P$_+$]{\includegraphics[width=8cm]{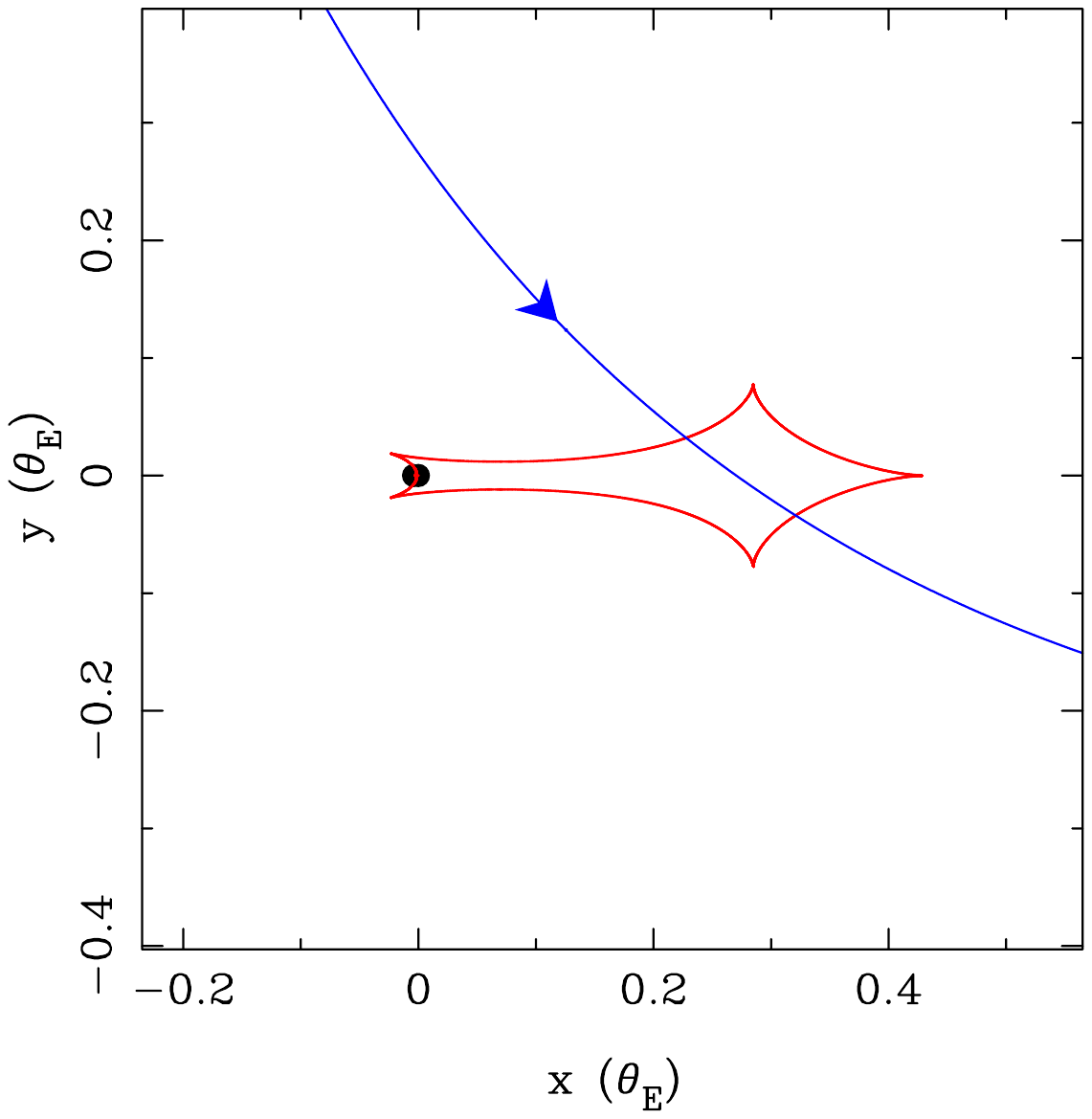}}
  \end{minipage}\hspace{-20pt}
  \begin{minipage}{0.35\textwidth}
    \centering
        \subfigure[Model-A P$_-$]{\includegraphics[width=8cm]{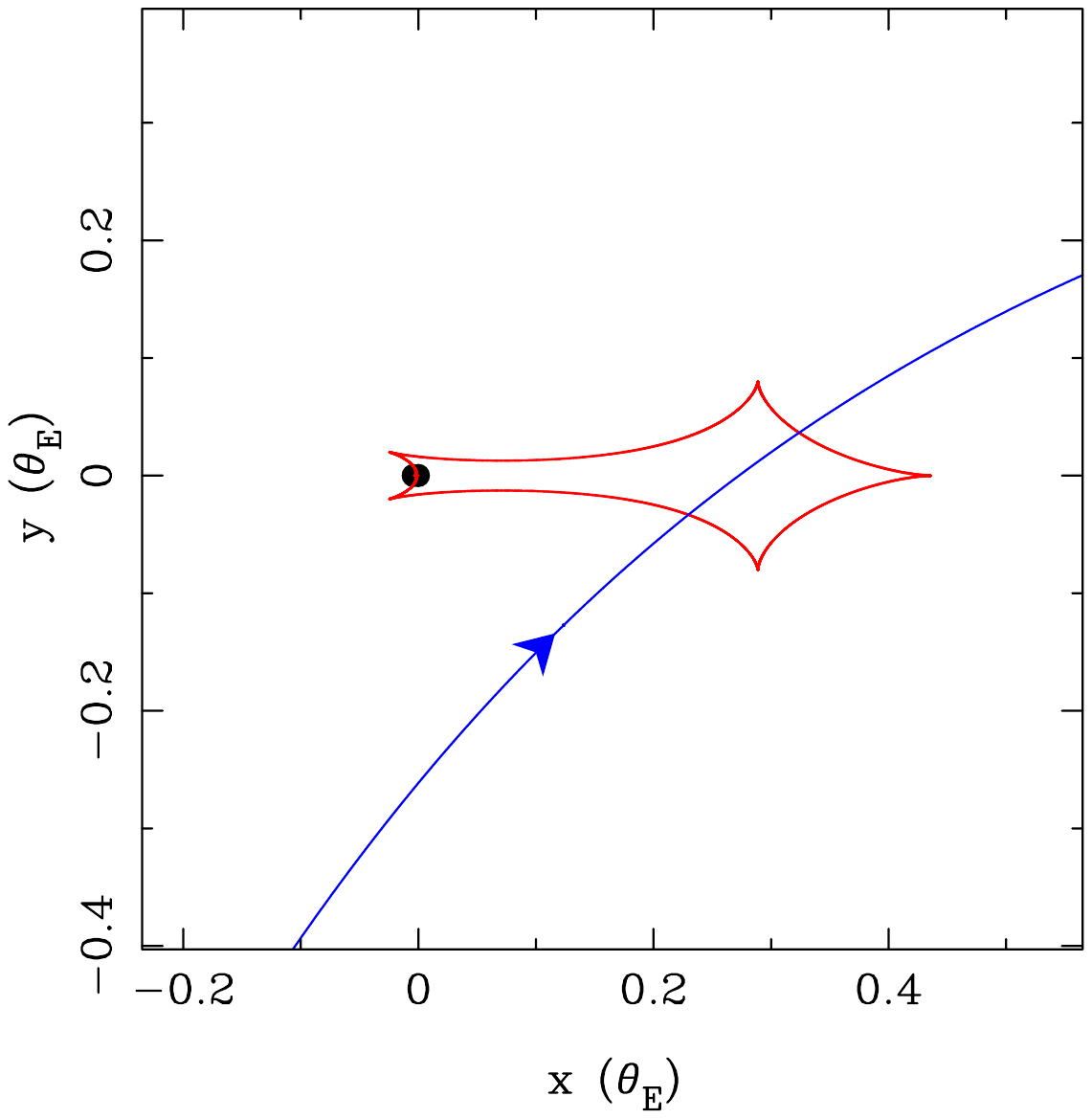}}
  \end{minipage}\hspace{-20pt}
  \caption{Caustic geometry corresponding to Model-A for versions with non-parallax and parallax (P$_\pm$). The caustic is shown in the red curved line. The blue line is the source trajectory. \label{fig:cauA}}
\end{figure*}
\begin{figure*}
  \hspace{-30pt}
  \begin{minipage}{0.35\textwidth}
    \centering
        \subfigure[Model-B]{\includegraphics[width=8cm]{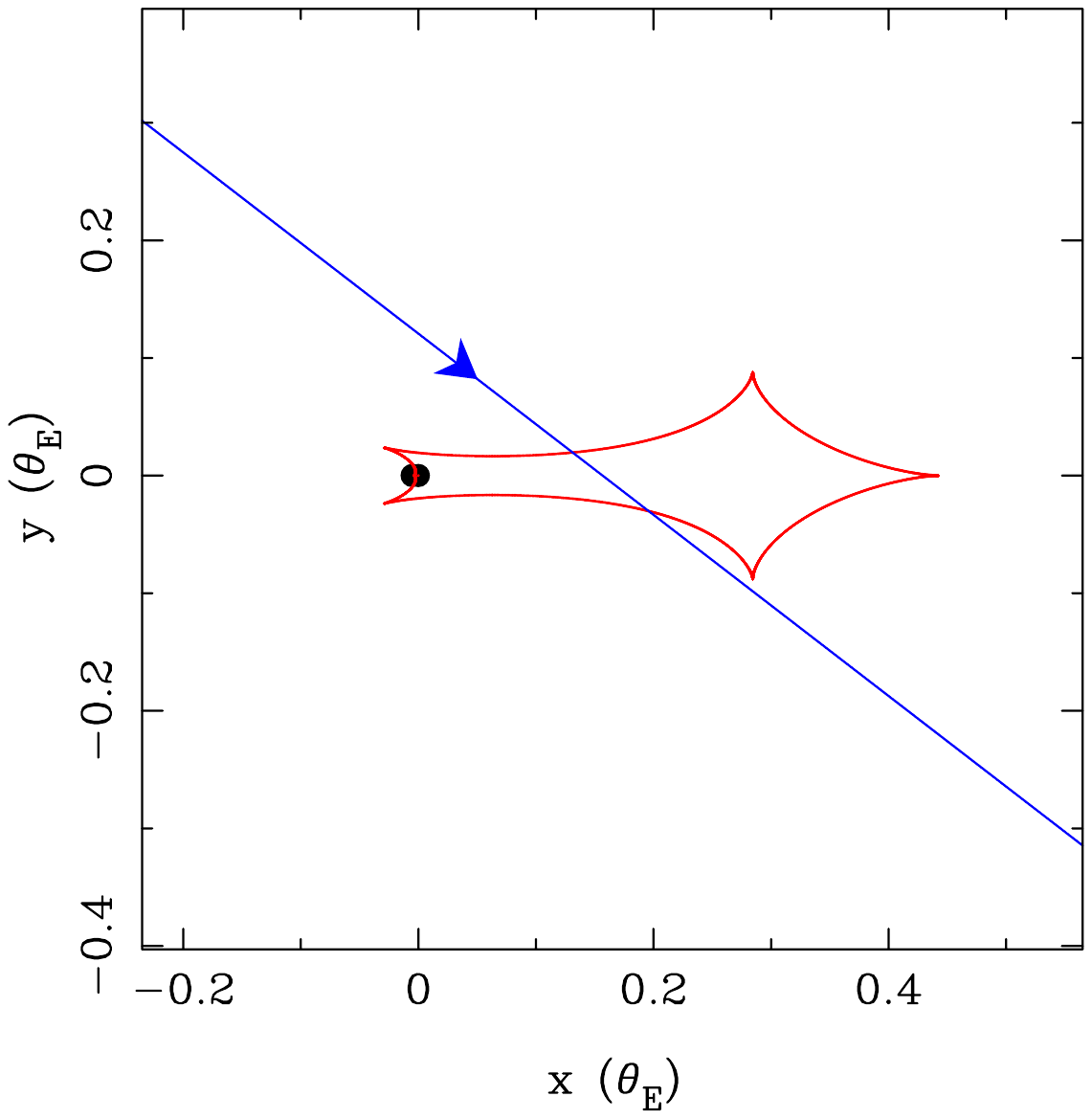}}
  \end{minipage}\hspace{-20pt}
  \begin{minipage}{0.35\textwidth}
    \centering
        \vspace{-10pt}
        \subfigure[Model-B P$_+$]{\includegraphics[width=8cm]{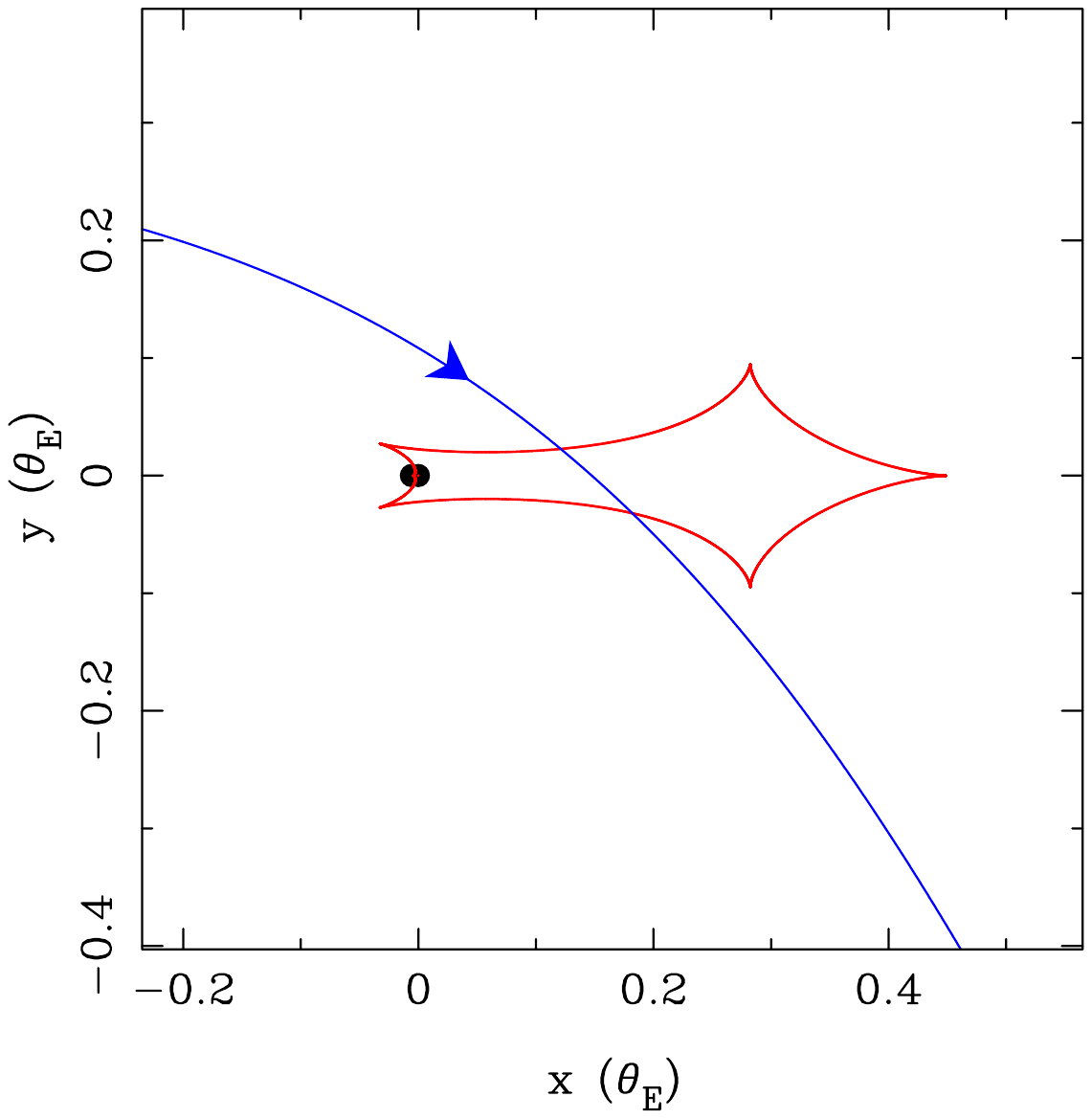}}
        
        \subfigure[Model-B P$_-$]{\includegraphics[width=8cm]{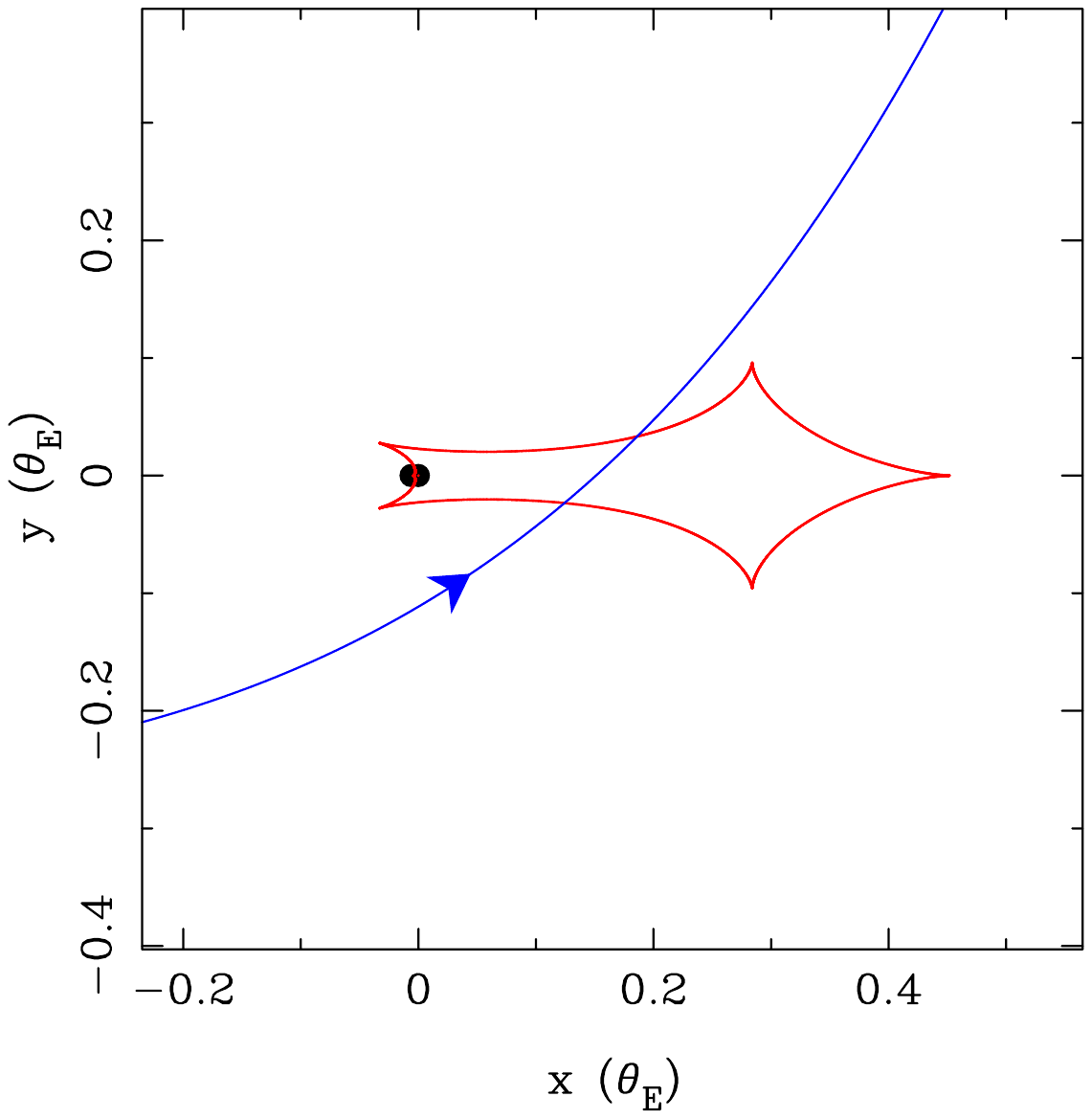}}
  \end{minipage}\hspace{-20pt}
  \begin{minipage}{0.35\textwidth}
    \centering
        \vspace{-10pt}
        \subfigure[Model-B P+LOM$_+$]{\includegraphics[width=8cm]{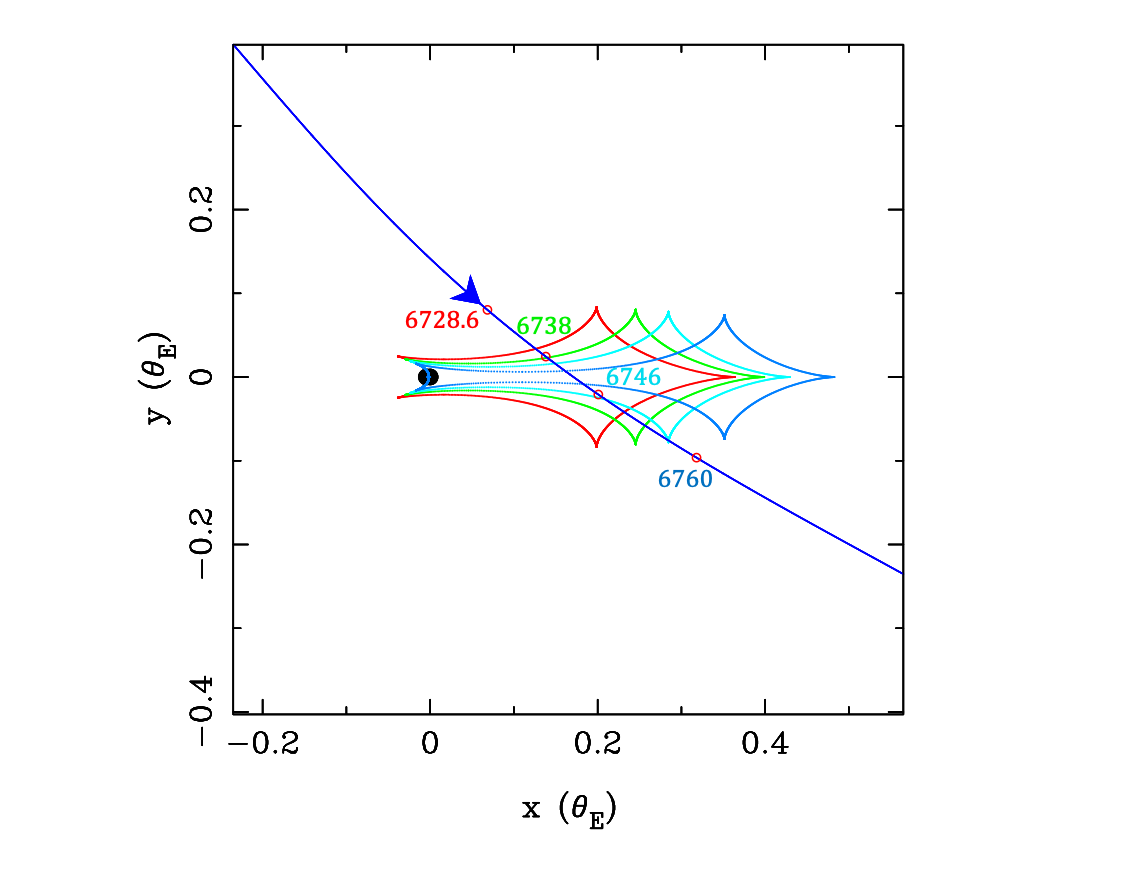}}
        
        \subfigure[Model-B P+LOM$_-$]{\includegraphics[width=8cm]{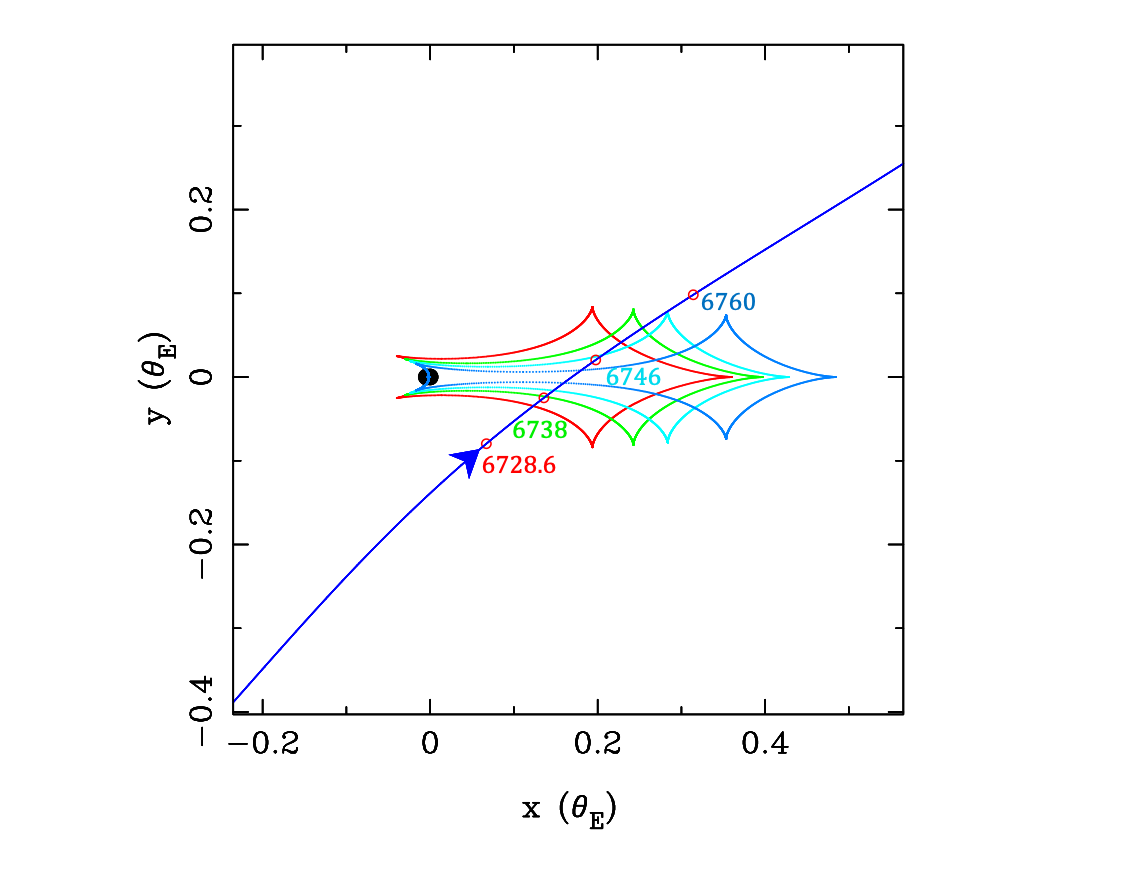}}
  \end{minipage}
  \caption{Caustic geometry corresponding to Model-B for versions with non-parallax, parallax (P$_\pm$) and parallax plus lens orbital motion (P+LOM$_\pm$). The caustics at $\mathrm{HJD'}=6728.6$, $6738$, $6746$ and $6760$ are plotted for P+LOM$_\pm$ with the instantaneous source positions in the red open circles whose sizes are not in scale. \label{fig:cauB}}
\end{figure*}

\section{Light Curve Analysis} \label{sec:modelling}
The light curve of OGLE-2014-BLG-0221 in Figure \ref{fig:1} shows a clear anomaly from $\mathrm{HJD'}$ $\sim$ 6738 to 6746, a deviation from the symmetric Paczynski curve \citep{Paczynski+1986} with a caustic crossing like shape, indicating the presence of a binary lens system.
We modeled the light curve with the following formularization.\par
The timescale of the event is defined as
\begin{eqnarray}
t_\mathrm{E} = \frac{\theta_\mathrm{E}}{\mu_\mathrm{rel}},
\end{eqnarray}
where $\mu_\mathrm{rel}$ is the relative source-lens proper motion and 
\begin{eqnarray}
\theta_\mathrm{E} = \sqrt{\kappa M \pi_\mathrm{rel}},
\end{eqnarray}
is the angular Einstein radius as a function of the constant $\kappa = 4G/c^2$, the total lens mass $M$, and the relative source-lens parallax $\pi_\mathrm{rel}=(1\ \mathrm{au})(1/D_\mathrm{L}-1/D_\mathrm{S})$.\par
The time at the source-lens closest approach and the impact parameter are parameterized as $t_0$ and $u_0$. 
The ratio of the angular source radius $\theta_\ast$ and the angular Einstein radius is also measured as 
\begin{eqnarray}
\rho = \frac{\theta_\ast}{\theta_\mathrm{E}}.
\end{eqnarray}

\vspace*{-\baselineskip}
\begin{deluxetable*}{cccc}
\tablecaption{Model-A Parameters\label{tab:1A}}
\tabletypesize{\scriptsize}
\tablehead{
\colhead{Parameter} &
\multicolumn{1}{c}{Model-A} & \multicolumn{1}{c}{Model-A P$_+$} & \multicolumn{1}{c}{Model-A P$_-$}
} 
\startdata
$t_\mathrm{E}$ (day) & \multicolumn{1}{c}{$67.6 \pm 1.3$} & \multicolumn{1}{c}{$73.1 \pm 3.0$} & $70.9 \pm 2.3$ \\
$t_0$ (HJD') & \multicolumn{1}{c}{$6728.34 \pm 0.14$} & \multicolumn{1}{c}{$6728.20 \pm 0.15$} & $6728.17 \pm 0.16$ \\
$u_0$ & \multicolumn{1}{c}{$0.1773 \pm 0.0038$} & \multicolumn{1}{c}{$0.1764 \pm 0.0047$} & $-0.1774 \pm 0.0039$ \\
$q$ $(10^{-3})$ & \multicolumn{1}{c}{$6.03 \pm 0.34$} & \multicolumn{1}{c}{$4.64 \pm 0.62$} & $5.02 \pm 0.54$ \\
$s$ & \multicolumn{1}{c}{$1.1517 \pm 0.0049$} & \multicolumn{1}{c}{$1.1516 \pm 0.0059$} & $1.1538 \pm 0.0048$ \\
$\alpha$ (rad) & \multicolumn{1}{c}{$0.701 \pm 0.012$} & \multicolumn{1}{c}{$0.793 \pm 0.030$} & $0.771 \pm 0.028$ \\
$\rho$ $(10^{-4})$ & \multicolumn{1}{c}{$7.80 \pm 1.54$} & \multicolumn{1}{c}{$6.94 \pm 1.76$} & $7.08 \pm 2.09$ \\
$\pi_\mathrm{E,N}$ & \multicolumn{1}{c}{---} & \multicolumn{1}{c}{$-0.556 \pm 0.164$} & $0.431 \pm 0.157$ \\
$\pi_\mathrm{E,E}$ & \multicolumn{1}{c}{---} & \multicolumn{1}{c}{$0.100 \pm 0.046$} & $0.093 \pm 0.045$ \\
$\pi_\mathrm{E}$ & \multicolumn{1}{c}{---} & \multicolumn{1}{c}{$0.565 \pm 0.167$} & $0.441 \pm 0.160$ \\
\hline
$\chi^2$ & \multicolumn{1}{c}{$10489.4$} & \multicolumn{1}{c}{$10483.5$} & $10486.9$ \\
$\Delta\chi^2$ & \multicolumn{1}{c}{---} & \multicolumn{1}{c}{$-5.9$} & $-2.5$ 
\enddata
\tablecomments{Best-fit values with uncertainties corresponding to the 68\% credible intervals of the MCMC posterior distributions.}
\end{deluxetable*}
\vspace*{-\baselineskip}
\vspace*{-\baselineskip}
\begin{deluxetable*}{cccccc}
\tablecaption{Model-B Parameters\label{tab:1B}}
\tabletypesize{\scriptsize}
\tablehead{
\colhead{Parameter} & 
\colhead{Model-B} & \multicolumn{1}{c}{Model-B P$_+$} & \multicolumn{1}{c}{Model-B P$_-$} & 
\colhead{Model-B P+LOM$_+$} & \colhead{Model-B P+LOM$_-$}
} 
\startdata
$t_\mathrm{E}$ (day) & $104.5 \pm 0.3$ & \multicolumn{1}{c}{$108.2 \pm 3.9$} & $105.3 \pm 3.5$ & $109.8 \pm 3.5$ & $112.9 \pm 2.4$\\
$t_0$ (HJD') & $6728.63 \pm 0.06$ & \multicolumn{1}{c}{$6728.92 \pm 0.13$} & \multicolumn{1}{c}{$6728.92 \pm 0.14$} & $6728.64 \pm 0.09$ & $6728.59 \pm 0.15$\\
$u_0$ & $0.0957 \pm 0.0003$ & \multicolumn{1}{c}{$0.0914 \pm 0.0025$} & \multicolumn{1}{c}{$-0.0941 \pm 0.0039$} & $0.1054 \pm 0.0031$ & $-0.1042 \pm 0.0015$\\
$q$ $(10^{-3})$ & $6.02 \pm 0.05$ & \multicolumn{1}{c}{$6.96 \pm 0.53$} & \multicolumn{1}{c}{$7.14 \pm 0.30$} & $4.73 \pm 0.16$ & $4.74 \pm 0.30$\\
$s$ & $1.1510 \pm 0.0007$ & \multicolumn{1}{c}{$1.1496 \pm 0.0040$} & \multicolumn{1}{c}{$1.1506 \pm 0.0024$} & $1.1026 \pm 0.0021$ & $1.1000 \pm 0.0045$\\
$\alpha$ (rad) & $0.656 \pm 0.003$ & \multicolumn{1}{c}{$0.598 \pm 0.022$} & \multicolumn{1}{c}{$-0.597 \pm 0.006$} & $0.705 \pm 0.004$ & $-0.700 \pm 0.013$\\
$\rho$ $(10^{-4})$ & $0.38 \pm 0.15$ & \multicolumn{1}{c}{$1.40 \pm 0.79$} & \multicolumn{1}{c}{$1.44 \pm 0.19$} & $0.98 \pm 0.17$ & $1.11 \pm 0.18$\\
$\pi_\mathrm{E,N}$ & --- & \multicolumn{1}{c}{$0.304 \pm 0.093$} & \multicolumn{1}{c}{$-0.337 \pm 0.022$} & $-0.419 \pm 0.010$ & $0.390 \pm 0.062$\\
$\pi_\mathrm{E,E}$ & --- & \multicolumn{1}{c}{$-0.070 \pm 0.032$} & \multicolumn{1}{c}{$-0.060 \pm 0.018$} & $-0.010 \pm 0.018$ & $-0.058 \pm 0.026$\\
$\pi_\mathrm{E}$ & --- & \multicolumn{1}{c}{$0.312 \pm 0.085$} & \multicolumn{1}{c}{$0.342 \pm 0.022$} & $0.419 \pm 0.009$ & $0.394 \pm 0.059$\\
$\gamma_\mathrm{\parallel}$ $(\mathrm{yr^{-1}})$ & --- & \multicolumn{1}{c}{---} & \multicolumn{1}{c}{---} & $1.35 \pm 0.03$ & $1.05 \pm 0.08$\\
$\gamma_\mathrm{\perp}$ (rad $\mathrm{yr^{-1}}$) & --- & \multicolumn{1}{c}{---} & \multicolumn{1}{c}{---} & $1.01 \pm 0.02$ & $-1.36 \pm 0.39$\\
\hline
$\chi^2$ & $10534.2$ & \multicolumn{1}{c}{$10475.3$} & \multicolumn{1}{c}{$10473.5$} & $10449.5$ & $10457.6$\\
$\Delta\chi^2$ & --- & \multicolumn{1}{c}{$-58.9$} & \multicolumn{1}{c}{$-60.7$} & $-84.7$ & $-76.6$\\
$(\mathrm{KE}/\mathrm{PE})_\mathrm{\perp}$ & --- & \multicolumn{1}{c}{---} & \multicolumn{1}{c}{---} & $0.373$ & $0.477$
\enddata
\tablecomments{Best-fit values with uncertainties corresponding to the 68\% credible intervals of the MCMC posterior distributions.}
\end{deluxetable*}
\vspace*{-\baselineskip}

For the case of a binary lens event, the mass ratio of the primary and companion $q$, the projected angular separation of them in units of the angular Einstein radius $s$, and the sky projected source incident angle $\alpha$ relative to the binary lens axis are included as additional parameters.
We considered annual parallax effect and linear lens orbital motion modeled by two additional parameters for each, ($\pi_\mathrm{E,N}$, $\pi_\mathrm{E,E}$) and ($\gamma_\mathrm{\parallel}=\frac{ds}{dt}$, $\gamma_\mathrm{\perp}=\frac{d\alpha}{dt}$), where $\pi_\mathrm{E,N}$ and $\pi_\mathrm{E,E}$ are the north and east components of the microlensing parallax vector \citep{Gould+2004} and $\frac{ds}{dt}$ and $\frac{d\alpha}{dt}$ are the projected linear and angular motions of the lens companion, expressed as the rate of change in $s$ and $\alpha$, around the primary \citep{Skowron+2011}.  \par

As boundary conditions of a gravitationally bound lens companion, we constrain the orbital motion parameters such that the ratio of the projected kinetic to potential energy of the lens $(\frac{\mathrm{KE}}{\mathrm{PE}})_\mathrm{\perp}$ does not go beyond 0.5, under an assumption of a circular orbit \citep{Dong+2009}.
Source orbital motion, another higher order effect called xallarap \citep{Poindexter+2005}, arising if the source star has a companion, is known to mimic the effects of annual parallax and lens orbital motion. While we also modeled this effect with seven additional parameters $(\xi_\mathrm{E,N}, \xi_\mathrm{E,E}, \mathrm{R.A.}_\xi, \mathrm{decl.}_\xi, P_\xi, e_\xi, T_\mathrm{peri})$, we will not go into detail in this paper since we found no solid evidence for xallarap. 
\par
We applied a linear limb darkening model to the source star in our modeling.
With $T_\mathrm{eff} \sim 5500$ K estimated from the source color (see Section \ref{sec:CMD}) and an assumption of surface gravity $\text{log}\ g=4.5\ \text{cm/s$^2$}$ and solar metallicity $\text{log}[M/H]=0$, the limb darkening coefficients $u_\mathrm{I}=0.5189$, $u_\mathrm{V}=0.6854$ and $u_\mathrm{MOA-Red}=0.6052$ were determined from the tables of \cite{Claret+2011}.

We searched for the best-fit binary lens model by exploring the parameter space using our modeling software \citep{sumi+2010} which is based on the Markov Chain Monte Carlo method \citep{Verde+2003} and the image centered ray-shooting method \citep{Bennett+1996}.
In order to explore a wide range of parameter values, we start from a grid search in which we divide the ($q, s, \alpha$) parameter space into $11 \times 22 \times 40$ uniformly spaced grids 
with $\mathrm{log}q\in[-4, 0], \mathrm{log}s\in[-0.5, 0.55], \alpha\in[0, 2\pi)$. We search for a best-fit model at each grid by fixing ($q, s, \alpha$) and allowing other parameters free. 
Among the 9680 models, 100 models with the lowest $\chi^2$ values are then refined to explore the global minimum by releasing the fixed ($q, s, \alpha$) to vary. We finally exclude all the models that exceed the threshold of $\Delta\chi^2=100$ from the best-fit lowest $\chi^2$ model, through which we find two degenerate models remain. \par

Figure \ref{fig:2} shows the two degenerate models, Model-A and Model-B, with corresponding geometries of the caustics shown in Figure \ref{fig:cauA} and \ref{fig:cauB}. The non-parallax, parallax and parallax plus lens orbital motion (LOM) models are plotted in Figure \ref{fig:2} and denoted by the unmarked model names, the extra character P and P+LOM, respectively. 
Subscripts ``$+$'' and ``$-$'' refer to degenerate parallax models with corresponding signs of the impact parameter $u_0$. 
Parameters of the models are listed in Tables \ref{tab:1A} and \ref{tab:1B}. \par

The event timescales of both Model-A and Model-B, $t_\mathrm{E}\sim70$ days and $110$ days respectively, are longer than the typical timescale of $t_\mathrm{E}\sim30$ days for events toward the Galactic bulge \citep{Mroz+2019}, from which we anticipate a heavy lens. The Model-A resembles the Model-B in the binary lens parameters $q, s$ and $\alpha$, but differs in $t_\mathrm{E}$ and $\rho$. The larger $t_\mathrm{E}$ value and the smaller $\rho$ value of Model-B implies a larger $\theta_\mathrm{E}$ compared to that of Model-A.
\par
The non-parallax model of Model-A already fits well to the light curve whereas Model-B shows a deviation around $\mathrm{HJD'} \sim 6760$ due to cusp re-approach after the caustic exit, resulting in $\Delta\chi^2\sim45$ between Model-A and Model-B. Once the higher order effects are considered, Model-B significantly improves from the non-parallax model by avoiding the cusp re-approach and $\Delta\chi^2\sim-60$ and $\sim-85$ for the parallax and parallax plus LOM models. In contrast, Model-A only improves by $\Delta\chi^2=-6$, showing no distinct evidence of the parallax signal. Therefore, we conclude a higher order effect of either parallax, lens orbital motion or xallarap needs to be involved in Model-B but not in Model-A.
\par
In order to take the higher order effect into account for Model-B, we begin with the parallax alone model that already provides a good fit to the flat feature around $\mathrm{HJD'} \sim 6760$ and then add LOM for getting more rigorous parallax parameter values and errors. Our modeling suffers a degeneracy known as the ecliptic degeneracy, where light curves with the parameters $(u_0, \alpha, \pi_\mathrm{E,N}, \gamma_\mathrm{\perp})=-(u_0, \alpha, \pi_\mathrm{E,N}, \gamma_\mathrm{\perp})$ appear nearly identical if direction of the Sun's acceleration is constant \citep{Skowron+2011}. This occurs when a deviation due to parallax is short timescale or the event coordinate is close to the ecliptic plane which is true for the Galactic bulge. We identify the degenerate models with the subscripts ``$+$'' and ``$-$'' taken from the sign of $u_0$. The ``$+$'' and ``$-$'' models have similar parameter values except the sign and are indeed indistinguishable from the light curve and $\chi^2$.
\par

\begin{figure}[htb!]
\includegraphics[width=\columnwidth]{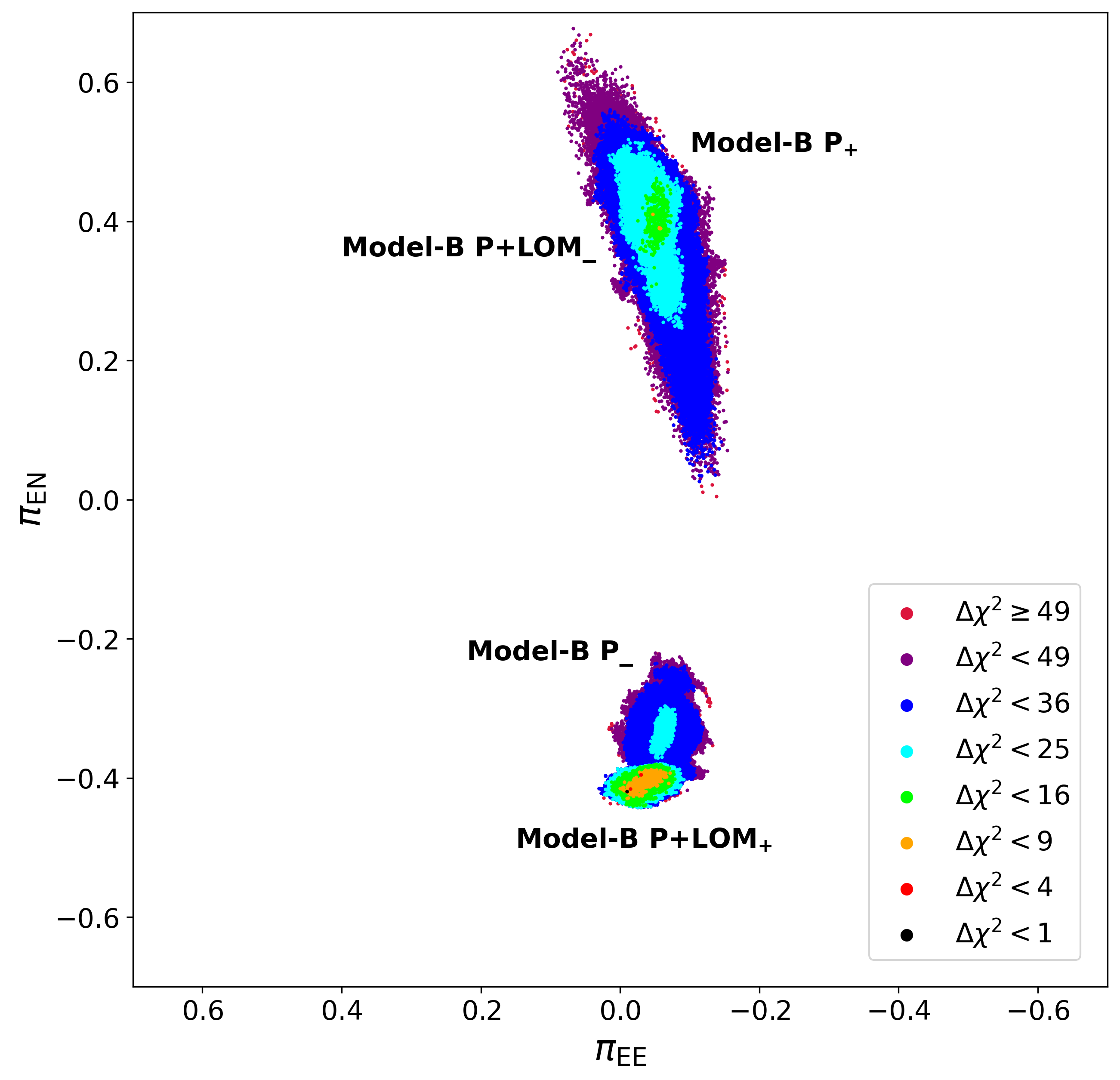}
\caption{Posterior distribution sampled by the MCMC sampler. Colors represent $\Delta\chi^2$ from the lowest $\chi^2$ parallax plus LOM model. \label{fig:2.5}}
\end{figure}

\begin{figure*}[htb!]
\gridline{\fig{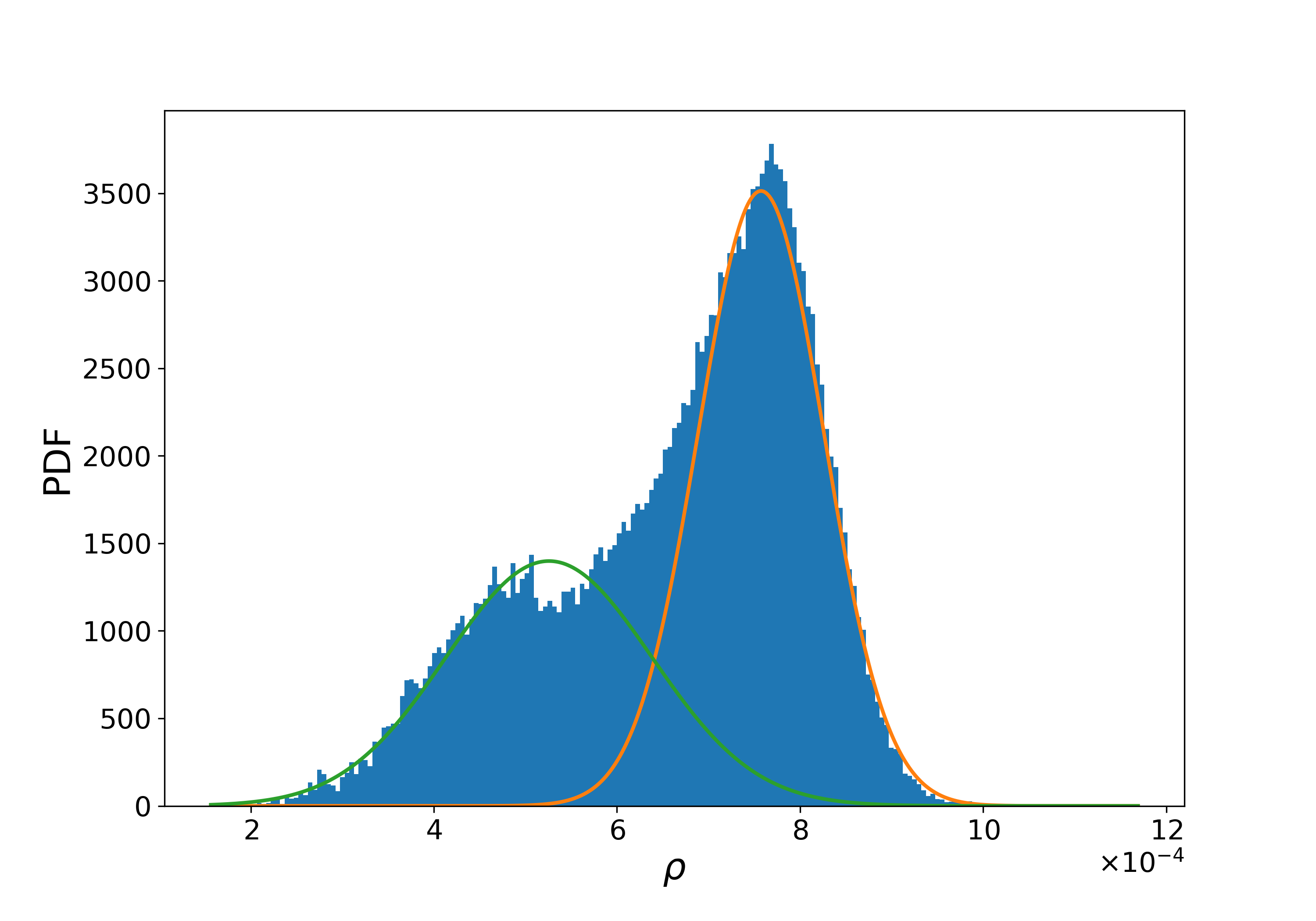}{0.35\textwidth}{(a)}
          \hspace{-20pt}
          \fig{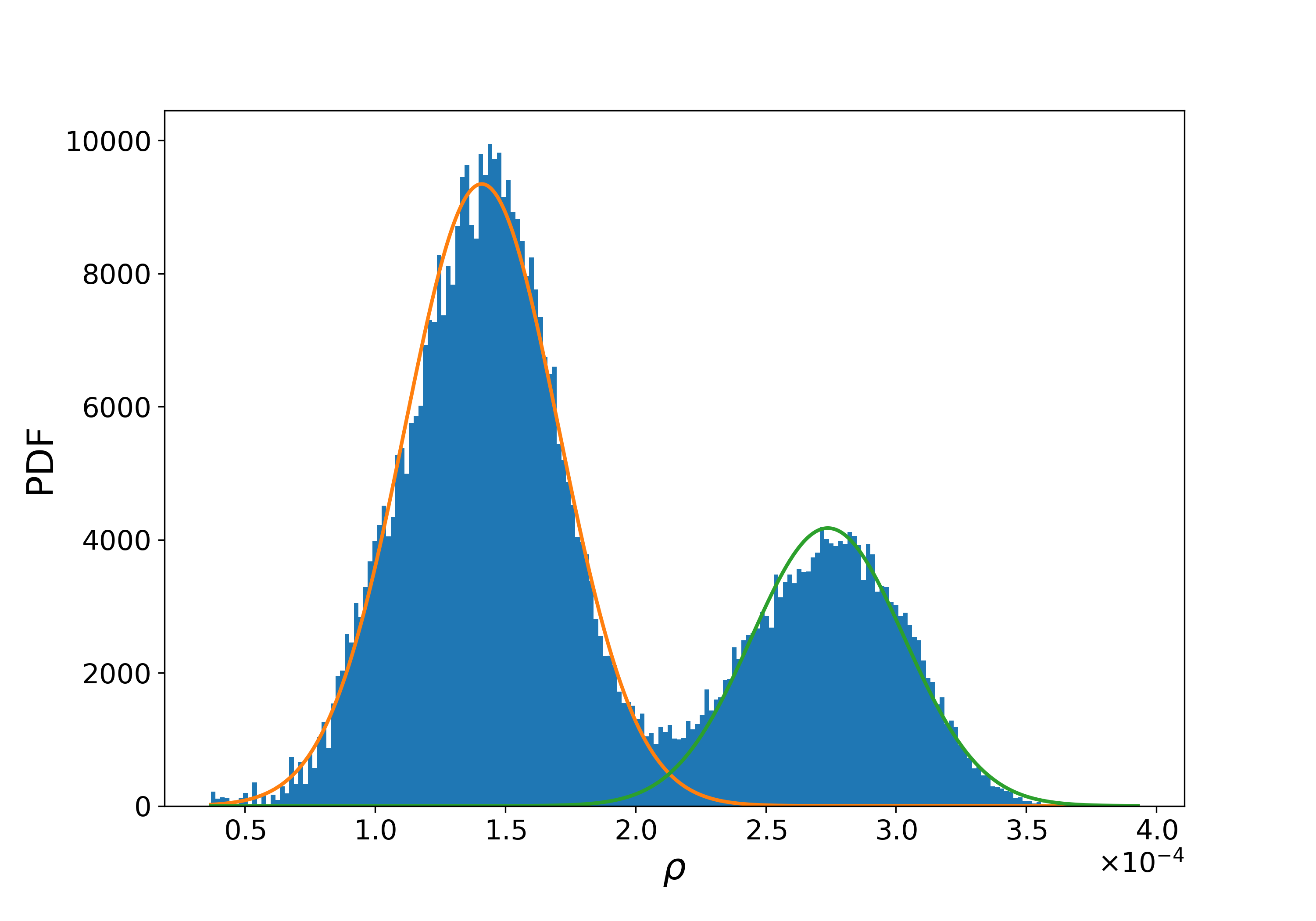}{0.35\textwidth}{(b)}
          \hspace{-20pt}
          \fig{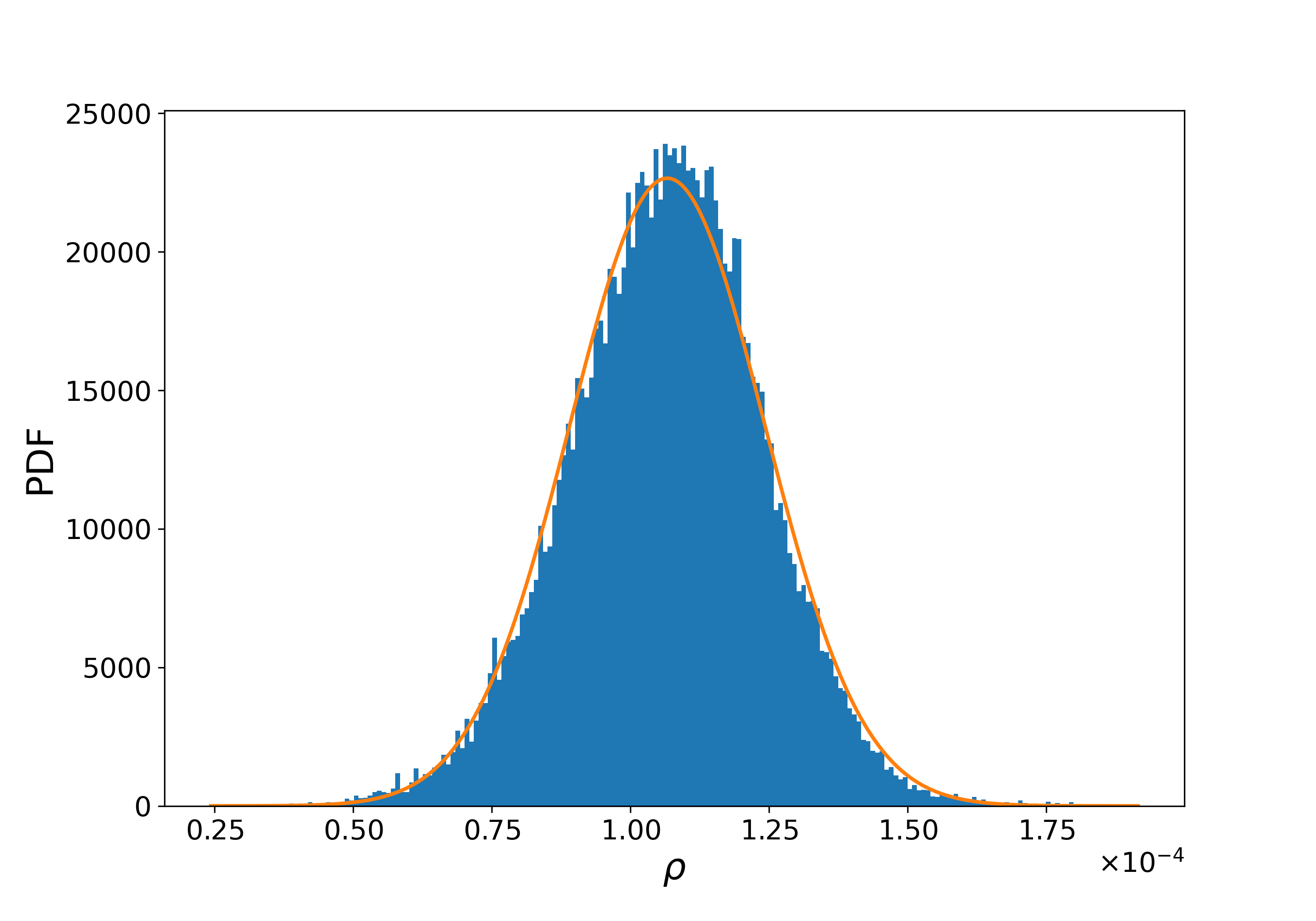}{0.35\textwidth}{(c)}
          }
\caption{Probability density functions (PDFs) of the normalized angular source radius $\rho$ from our MCMC chains for (a) Model-A, (b) Model-B P$_+$ and (c) Model-B P+LOM$_+$. The bimodal distribution of each model except Model-B P+LOM$_+$ is approximated by the Gaussian Mixture Model, and two components of the best fit is plotted in orange and green lines. \label{fig:3}}
\end{figure*}
\vspace*{-\baselineskip}

The inclusion of LOM improves the fitting by $\Delta\chi^2\sim-25$ and reduces uncertainty of the parallax vector from $\sim30\%$ to $\sim15\%$ in the positive $\pi_\mathrm{E,N}$ regime and $\sim6\%$ to $\sim2\%$ in the negative $\pi_\mathrm{E,N}$ regime, as can be noticed in Tables \ref{tab:1A} and \ref{tab:1B} and Figure \ref{fig:2.5} where we show the posterior distribution of the parallax vector components with uniform prior of $\pi_\mathrm{E,N}$ and $\pi_\mathrm{E,E}$. However, we find the $\Delta\chi^2$ mostly comes from the baseline systematics but not from the magnification part of the light curve, which indicates that an effective $\chi^2$ improvement due to the microlensing signal is negligibly small. We also examined the xallarap effect by simultaneously fitting with the parallax parameters, but it changes neither the model nor the parallax vector much. We note that values of $t_\mathrm{E}$ and $\rho$ are consistent between the parallax alone model and the parallax plus LOM/xallrap model.\par

Despite the strict constraint of the parallax parameters imposed by simultaneous fitting of the parallax effect and LOM, we confirmed that the deviation around $\mathrm{HJD'} \sim 6760$ is also explainable by the LOM or xallarap alone, which accepts null detection of the parallax. We thus conclude the parallax parameters cannot be determined contrary to the small model uncertainty.

\par

There is also a large dispersion in the distribution of $\rho$ as shown in Figure \ref{fig:3} where the posterior distributions of $\rho$ are plotted. The dispersion is due to the sparsity of data points taken during the caustic entry and exit. The distributions are well characterized by the superposition of two Gaussian distributions, and the best-fit finite source parameter of the models lies either of the bimodal peaks. This bimodal feature is associated with the uncertainty of $\rho$, causing it to be larger.

\begin{figure}[ht!]
\includegraphics[width=\linewidth]{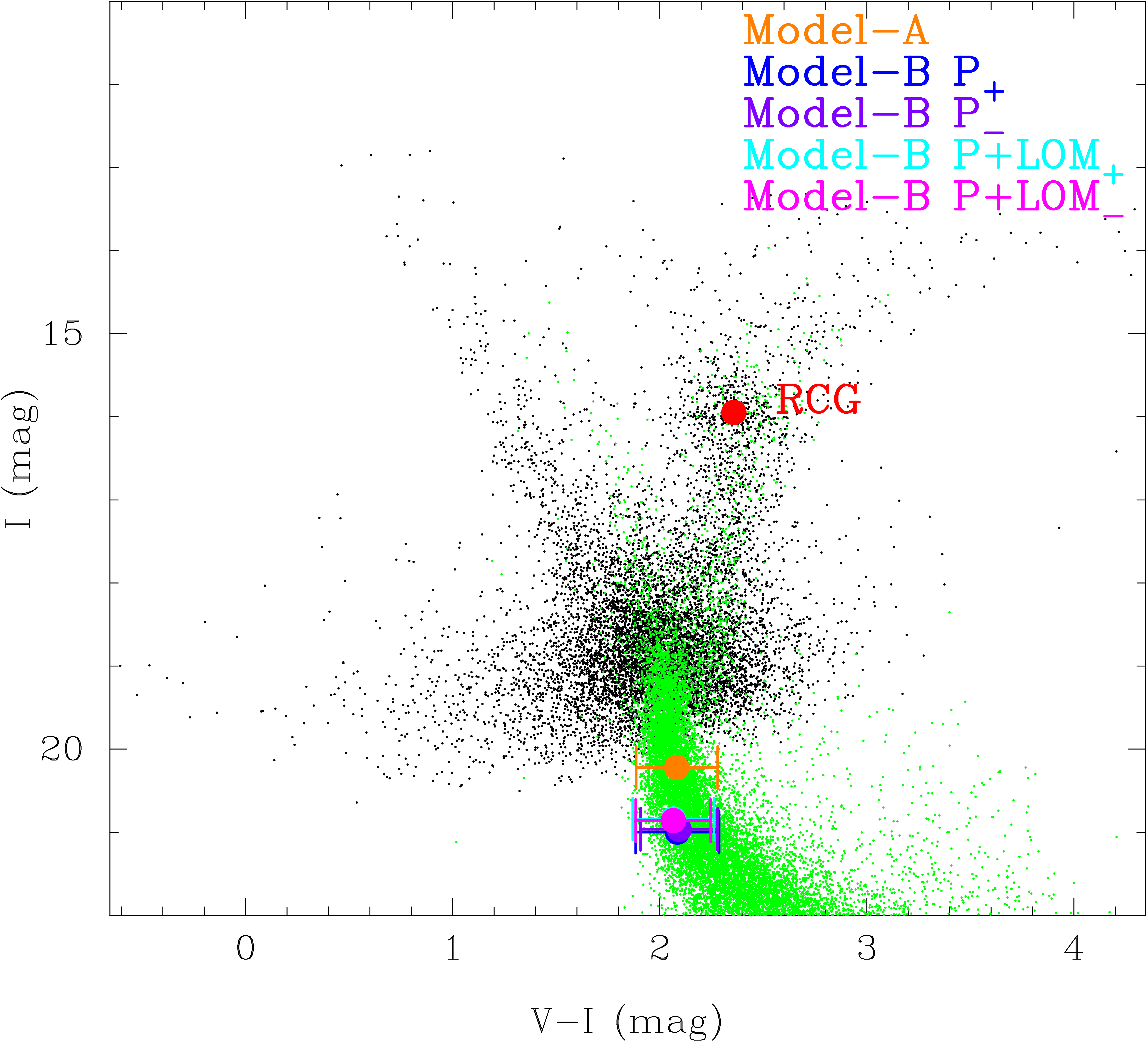}
\caption{$(V-I, I)$ Color-Magnitude Diagram of stars in the OGLE-I\hspace{-.1em}I\hspace{-.1em}I catalog \citep{szymanski+2011} and the Hubble Space Telescope catalog \citep{Holtzman+1998}. The black dots are the OGLE-I\hspace{-.1em}I\hspace{-.1em}I catalog stars located within $2'$ of the event OGLE-2014-BLG-0221, and the green dots are the Hubble Space Telescope catalog stars in a field of Baade's Window, whose colors and magnitudes are shifted to match the red clump giant centroid of the OGLE-I\hspace{-.1em}I\hspace{-.1em}I catalog. The red spot indicates position of the red clump giant centroid, and the orange, blue, purple, cyan and magenta spots are source positions of Model-A, Model-B P$\pm$ and Model-B P+LOM$\pm$, respectively. \label{fig:4}}
\end{figure}

\vspace*{-\baselineskip}
\begin{deluxetable*}{cccccc}
\tablecaption{Source/Lens Properties\label{tab:2}}
\tabletypesize{\scriptsize}
\tablewidth{0pt} 
\tabletypesize{\scriptsize}
\tablehead{
\colhead{} &
\multicolumn{1}{c}{Model-A} & \multicolumn{1}{c}{Model-B P$_+$} & \multicolumn{1}{c}{Model-B P$_-$} & 
\multicolumn{1}{c}{Model-B P+LOM$_+$} & 
\multicolumn{1}{c}{Model-B P+LOM$_-$}
} 
\startdata
\begin{tabular}{c}$(V-I, I)_\mathrm{S}$ \\ \ \end{tabular} & \multicolumn{1}{c}{\begin{tabular}{c}$(2.082, 20.221)$ \\ $\pm(0.183, 0.028)$\end{tabular}} & \multicolumn{1}{c}{\begin{tabular}{c}$(2.085, 20.998)$ \\ $\pm(0.183, 0.033)$\end{tabular}} & \multicolumn{1}{c}{\begin{tabular}{c}$(2.093, 20.964)$ \\ $\pm(0.185, 0.046)$\end{tabular}} & \begin{tabular}{c} $(2.066, 20.842)$ \\ $\pm(0.182, 0.035)$\end{tabular} & \begin{tabular}{c}$(2.065, 20.861)$ \\ $\pm(0.181, 0.017)$\end{tabular} \\
\begin{tabular}{c}$(V-I, I)_\mathrm{S,0}$ \\ \  \end{tabular} & \multicolumn{1}{c}{\begin{tabular}{c}$(0.784, 18.623)$ \\ $\pm(0.196, 0.054)$\end{tabular}} & \multicolumn{1}{c}{\begin{tabular}{c}$(0.787, 19.400)$ \\ $\pm(0.196, 0.057)$\end{tabular}} & \multicolumn{1}{c}{\begin{tabular}{c}$(0.795, 19.367)$ \\ $\pm(0.198, 0.065)$\end{tabular}} & \begin{tabular}{c}$(0.768, 19.244)$ \\ $\pm(0.195, 0.058)$\end{tabular} & \begin{tabular}{c}$(0.767, 19.263)$ \\ $\pm(0.194, 0.049)$\end{tabular} \\
$\theta_\ast$ ($\mu$as) & \multicolumn{1}{c}{$0.638\pm0.122$} & \multicolumn{1}{c}{$0.447\pm0.085$} & \multicolumn{1}{c}{$0.457\pm0.089$} & $0.471\pm0.090$ & $0.467\pm0.088$ \\
$\theta_\mathrm{E}$ (mas) & $0.82\pm0.23$ & $3.19\pm1.86$ & $3.19\pm0.76$ & $4.83\pm1.25$ & $4.20\pm1.05$ \\
$\mu_\mathrm{rel}$ (mas/year) & $4.42\pm1.22$ & $10.77\pm6.29$ & $11.07\pm2.65$ & $16.08\pm4.19$ & $13.59\pm3.42$ \\
\enddata
\end{deluxetable*}
\vspace*{-\baselineskip}

\section{Source Property}
\label{sec:CMD} 
Once the finite source effect is detected, the angular Einstein radius $\theta_\mathrm{E}=\frac{\theta_\ast}{\rho}$ can be determined by measuring the angular size of the source, and the relative proper motion $\mu_\mathrm{rel}=\frac{\theta_\mathrm{E}}{t_\mathrm{E}}$ can also be estimated, combining with $t_\mathrm{E}$ measured from the light curve. An empirical relation of the angular source size as a function of the cousins \textit{I} and \textit{V} band is derived based on the result of \cite{Boyajian+2014}. By restricting to stars with $3900\ \mathrm{K}<T_\mathrm{eff}<7000\ \mathrm{K}$, optimal to FGK stars, \cite{Fukui+2015} finds:
\begin{eqnarray}
\text{log}[2\theta_\ast/\mathrm{mas}]=0.5014+0.4197(V-I)-0.2I.
\end{eqnarray}
\par
We calibrated the source magnitude in the instrumental OGLE \textit{I}- and \textit{V}-band to the standard Kron-Cousins and Johnson system using the following equations from \cite{Udalski+2015}:
\begin{eqnarray}
(V-I) &=& \mu\cdot[(v_{DB}+\Delta ZP_V)-(i_{DB}+\Delta ZP_I)] \nonumber \\
I &=& (i_{DB}+\Delta ZP_I) + \epsilon_I\cdot(V-I) \\
V &=& (v_{DB}+\Delta ZP_V) + \epsilon_V\cdot(V-I), \nonumber
\end{eqnarray}
where $i_{DB}$ and $v_{DB}$ are the instrumental $I$ and $V$, $\mu=(1-\epsilon_V+\epsilon_I)^{-1}$, $\Delta ZP_I=0.018$, $\Delta ZP_V=0.158$, $\epsilon_I=-0.005\pm0.003$ and $\epsilon_V=-0.077\pm0.001$ for the field of OGLE-2014-BLG-0221.
The calibrated color and magnitude $(V-I, I)_\mathrm{S}$ are plotted onto the color-magnitude diagram (CMD) of the stars within $2'$ of the event coordinate (Figure \ref{fig:4}). \par
The central color and magnitude of the red clump giants (RCGs) population marked in the CMD is representative of the bulge RCGs. We followed the standard procedure adopted in \cite{Yoo+2004} to correct the effect of reddening and extinction due to interstellar dust, under the assumption that the source experiences the same reddening and extinction as the bulge RCGs. 
The centroid of the RCGs was $(V-I, I)_\mathrm{RCG}=(2.358, 15.947)\pm(0.009, 0.023)$ from the color-magnitude distribution, and the extinction-free color and magnitude of the bulge RCGs were known as $(V-I, I)_\mathrm{RCG, 0}=(1.060, 14.349)\pm(0.070, 0.040)$ toward the event coordinate \citep{Bensby+2011, Nataf+2013}, and for which the reddening and extinction were estimated as $(E(V-I), A(I))=(1.298, 1.598)\pm(0.071, 0.046)$. Applying the same reddening and extinction and using the earlier empirical relation, we found the intrinsic color and magnitude of the source $(V-I, I)_\mathrm{S,0}$ and the angular source size $\theta_\ast$ for Model-A, Model-B P$\pm$ and Model-B P+LOM$\pm$, that let us calculate $\theta_\mathrm{E}$ and $\mu_\mathrm{rel}$. Those values are given in Table \ref{tab:2}.
\begin{figure*}
\centering
\includegraphics[width=17cm]{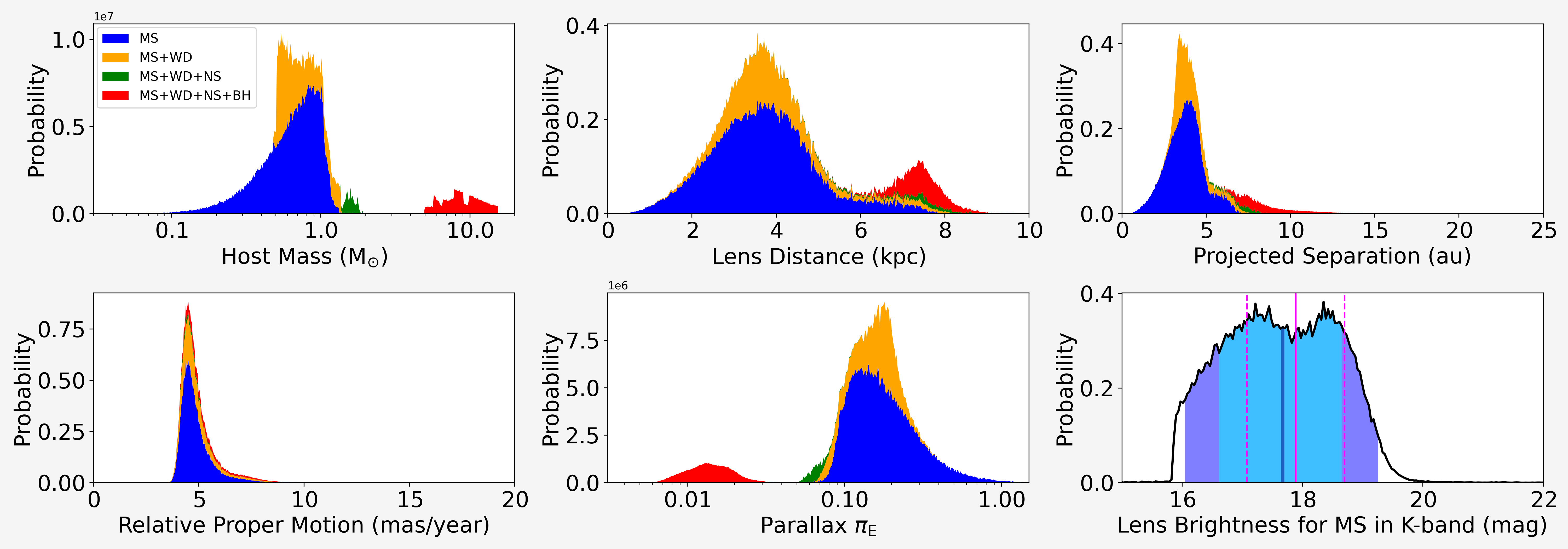}
\caption{Posterior probability density distributions of Model-A. The distributions are classified into main sequence stars, white dwarfs, neutrons stars and black holes and separated by colors accordingly. In the last plot, lens brightness of the main sequence samples is shown where the vertical blue line is the median value, and the cyan and purple regions represent 68.3 and 95.4\% credible interval of the distribution. The apparent source brightness is also plotted by the median magenta line with the magenta dashed line indicating the 68.3\% interval.}
\label{fig:5}
\end{figure*}
\begin{figure*}
\centering
\includegraphics[width=17cm]{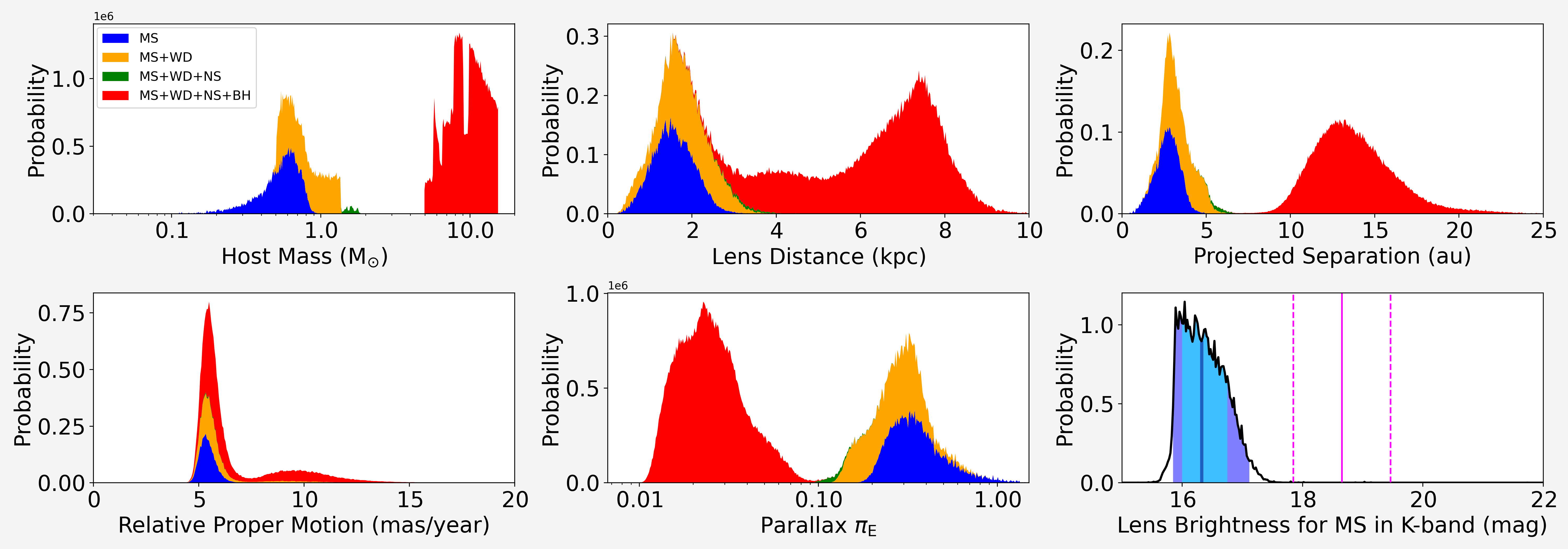}
\caption{Same as Figure \ref{fig:5} for Model-B P$_+$.}
\label{fig:7}
\end{figure*}

\begin{figure*}
\centering
\includegraphics[width=17cm]{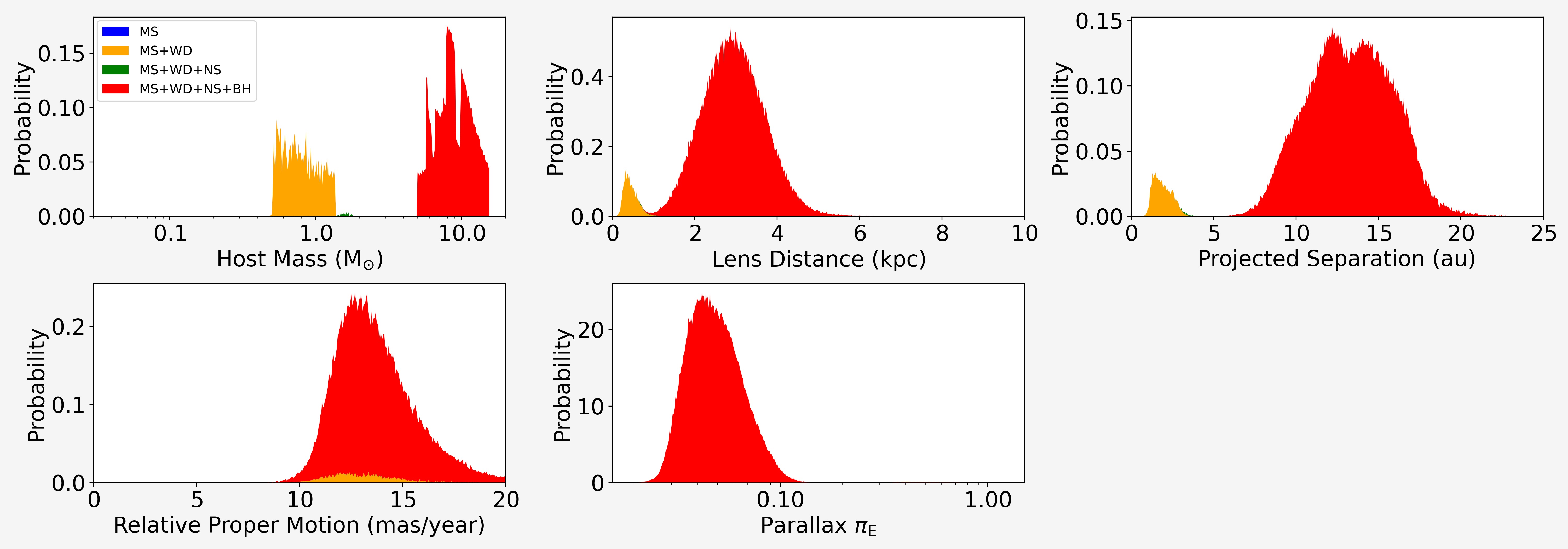}
\caption{Same as Figure \ref{fig:5} for Model-B P+LOM$_+$. The plot of the lens brightness is not shown as no main sequence sample has been generated in the simulation.}
\label{fig:9}
\end{figure*}
%
\vspace*{-\baselineskip}
\begin{deluxetable*}{ccccc}
\tablecaption{Estimated Lens Physical Parameters from the Bayesian Analysis\label{tab:3}}
\tabletypesize{\scriptsize}
\tablewidth{0pt} 
\tabletypesize{\scriptsize}
\tablehead{
\colhead{} & \colhead{} &
\multicolumn{1}{c}{Model-A} & \multicolumn{1}{c}{Model-B P$_+$} & 
\multicolumn{1}{c}{Model-B P+LOM$_+$}
} 
\startdata
\textbf{Main Sequence Host} & Companion Mass $M_\mathrm{comp}$ $(M_\mathrm{jup})$ & $4.35^{+1.65}_{-1.91}$ & $4.13^{+1.04}_{-1.28}$ &  --- \\
& Host Mass $M_\mathrm{host}$ $(M_\odot)$ & $0.69^{+0.26}_{-0.30}$ &  $0.57^{+0.14}_{-0.18}$ & --- \\
& Projected Separation $a_\mathrm{\perp}$ (au) & $3.76^{+0.96}_{-1.07}$ &  $2.75^{+0.65}_{-0.76}$ & --- \\
& Lens Distance $D_\mathrm{L}$ (kpc) & $3.68^{+1.13}_{-1.16}$ & $1.56^{+0.51}_{-0.50}$ & --- \\
& Source Distance $D_\mathrm{S}$ (kpc) & $7.97^{+2.51}_{-0.70}$ &  $7.87^{+1.27}_{-0.71}$ & --- \\
& Angular Einstein Radius $\theta_\mathrm{E}$ (mas) & $0.87^{+0.14}_{-0.08}$ & $1.52^{+0.13}_{-0.11}$ & --- \\
& Relative Proper Motion $\mu_\mathrm{rel}$ (mas/year) & $4.67^{+0.76}_{-0.37}$ & $5.41^{+0.43}_{-0.30}$ & --- \\
& Parallax $\pi_\mathrm{E}$ & $0.171^{+0.110}_{-0.048}$ & $0.351^{+0.156}_{-0.093}$ & --- \\
& Lens Brightness in \textit{V}-band $V_\mathrm{L}$ (mag) & $22.9^{+2.2}_{-1.8}$ & $21.5^{+0.8}_{-0.9}$ & --- \\
& Lens Brightness in \textit{I}-band $I_\mathrm{L}$ (mag) & $20.4^{+1.4}_{-1.3}$ & $18.9^{+0.6}_{-0.5}$ & --- \\
& Lens Brightness in \textit{H}-band $H_\mathrm{L}$ (mag) & $17.9^{+1.1}_{-1.1}$ & $16.7^{+0.4}_{-0.4}$ & --- \\
& Lens Brightness in \textit{K}-band $K_\mathrm{L}$ (mag) & $17.7^{+1.0}_{-1.1}$ & $16.3^{+0.4}_{-0.3}$ & --- \\
& Source Brightness in \textit{H}-band $H_\mathrm{S}$ (mag) & $18.1 \pm 0.8$ & $18.9 \pm 0.8$ & $18.8 \pm 1.2$ \\
& Source Brightness in \textit{K}-band $K_\mathrm{S}$ (mag) & $17.9 \pm 0.8$ & $18.7 \pm 0.8$ & $18.6 \pm 1.2$ \\
& Fraction of the Sample & 64.8\% & 18.0\% & --- \\
\hline
\textbf{White Dwarf Host} & Companion Mass $M_\mathrm{comp}$ $(M_\mathrm{jup})$ & $4.23^{+2.30}_{-0.76}$ & $5.68^{+2.51}_{-1.50}$ & $4.29^{+1.61}_{-1.27}$ \\
& Host Mass $M_\mathrm{host}$ $(M_\odot)$ & $0.67^{+0.36}_{-0.12}$ & $0.78^{+0.34}_{-0.21}$ & $0.87^{+0.32}_{-0.26}$ \\
& Projected Separation $a_\mathrm{\perp}$ (au)  & $3.83^{+1.05}_{-0.49}$ & $3.34^{+1.21}_{-0.71}$ & $1.87^{+0.68}_{-0.48}$ \\
& Lens Distance $D_\mathrm{L}$ (kpc) & $3.75^{+1.19}_{-0.76}$ & $1.85^{+0.69}_{-0.53}$ & $0.42^{+0.21}_{-0.11}$ \\
& Source Distance $D_\mathrm{S}$ (kpc) & $8.04^{+3.41}_{-0.79}$ & $7.84^{+1.12}_{-0.70}$ & $7.78^{+1.03}_{-0.80}$ \\
& Angular Einstein Radius $\theta_\mathrm{E}$ (mas) & $0.87^{+0.16}_{-0.08}$ & $1.60^{+0.28}_{-0.13}$ & $3.95^{+0.57}_{-0.47}$ \\
& Relative Proper Motion $\mu_\mathrm{rel}$ (mas/year) & $4.69^{+0.89}_{-0.40}$ & $5.64^{+0.69}_{-0.43}$ & $12.91^{+1.94}_{-1.39}$ \\
& Parallax $\pi_\mathrm{E}$ & $0.171^{+0.034}_{-0.070}$ & $0.266^{+0.103}_{-0.090}$ & $0.576^{+0.226}_{-0.164}$ \\
& Fraction of the Sample & 24.7\% & 20.8\% & 4.3\% \\
\hline
\textbf{Neutron Star Host} & Companion Mass $M_\mathrm{comp}$ $(M_\mathrm{jup})$ & $10.15^{+0.73}_{-0.82}$ & $11.75^{+0.90}_{-1.15}$ & $7.89^{+0.76}_{-0.45}$ \\
& Host Mass $M_\mathrm{host}$ $(M_\odot)$ & $1.61^{+0.11}_{-0.13}$ & $1.61^{+0.12}_{-0.16}$ & $1.59^{+0.15}_{-0.09}$ \\
& Projected Separation $a_\mathrm{\perp}$ (au) & $7.12^{+0.68}_{-1.46}$ & $5.71^{+0.50}_{-0.53}$ & $3.38^{+0.45}_{-0.29}$ \\
& Lens Distance $D_\mathrm{L}$ (kpc) & $7.23^{+0.66}_{-1.75}$ & $3.15^{+0.44}_{-0.60}$ & $0.77^{+0.29}_{-0.11}$ \\
& Source Distance $D_\mathrm{S}$ (kpc) & $12.20^{+2.04}_{-3.89}$ & $8.02^{+1.79}_{-0.72}$ & $7.81^{+0.98}_{-0.61}$ \\
& Angular Einstein Radius $\theta_\mathrm{E}$ (mas) & $0.86^{+0.10}_{-0.07}$ & $1.60^{+0.22}_{-0.14}$ & $3.86^{+0.45}_{-0.51}$ \\
& Relative Proper Motion $\mu_\mathrm{rel}$ (mas/year) & $4.61^{+0.54}_{-0.34}$ & $5.63^{+0.67}_{-0.44}$ & $12.61^{+1.49}_{-1.51}$ \\
& Parallax $\pi_\mathrm{E}$ & $0.065^{+0.013}_{-0.008}$ & $0.121^{+0.030}_{-0.002}$ & $0.051^{+0.031}_{-0.003}$ \\
& Fraction of the Sample & 2.0\% & 0.6\% & 0.1\% \\
\hline
\textbf{Black Hole Host} & Companion Mass $M_\mathrm{comp}$ $(M_\mathrm{jup})$ & $55.00^{+22.73}_{-14.91}$ & $72.16^{+24.50}_{-21.84}$ & $48.62^{+15.60}_{-13.62}$ \\
& Host Mass $M_\mathrm{host}$ $(M_\odot)$ & $8.71^{+3.60}_{-2.36}$ & $9.90^{+3.36}_{-3.00}$ & $9.81^{+3.15}_{-2.75}$ \\
& Projected Separation $a_\mathrm{\perp}$ (au) & $8.17^{+2.37}_{-1.04}$ &  $13.52^{+2.54}_{-1.99}$ & $13.30^{+2.69}_{-2.61}$ \\
& Lens Distance $D_\mathrm{L}$ (kpc) & $7.45^{+0.52}_{-0.52}$ & $6.75^{+1.04}_{-2.45}$ & $2.94^{+0.78}_{-0.75}$ \\
& Source Distance $D_\mathrm{S}$ (kpc) & $8.30^{+0.81}_{-0.59}$ & $8.87^{+2.69}_{-1.18}$ & $7.97^{+1.14}_{-0.65}$ \\
& Angular Einstein Radius $\theta_\mathrm{E}$ (mas) & $0.94^{+0.32}_{-0.12}$ & $1.73^{+1.18}_{-0.22}$ & $4.10^{+0.65}_{-0.47}$ \\
& Relative Proper Motion $\mu_\mathrm{rel}$ (mas/year) & $5.05^{+1.75}_{-0.59}$ & $5.95^{+3.94}_{-0.58}$ & $13.48^{+2.16}_{-1.57}$ \\
& Parallax $\pi_\mathrm{E}$ & $0.015^{+0.016}_{-0.006}$ & $0.025^{+0.021}_{-0.008}$ & $0.051^{+0.031}_{-0.003}$ \\
& Fraction of the Sample & 8.3\% & 60.6\% & 95.6\% \\
\enddata
\end{deluxetable*}
\vspace*{-\baselineskip}
\section{Lens Property} \label{sec:Bayesian}
Since parallax could not be measured accurately, we used a Bayesian approach to infer the lens physical properties. The microlensing event simulation code \citep{Koshimoto+2021a} with a parametric Galactic model toward the Galactic bulge developed by \cite{Koshimoto+2021b} was used to generate artificial microlensing events and obtain posterior distributions of the physical parameters. 
Likelihood of the measured parameters, $t_\mathrm{E}$ and $\theta_\mathrm{E}$, was included as the observed constraints. An upper limit of the lens brightness was also set as $I_\mathrm{lim}=18$ mag, determined based on the blending magnitude but about 2 mag brighter to make the analysis conservative. 
From the OGLE-I\hspace{-.1em}I\hspace{-.1em}I catalog, we also confirmed there is no potential source within $2''$ of the event coordinate that is brighter than the limit. Apparent lens magnitudes of the simulated events were estimated using the mass-luminosity and color-color relations of main sequence stars \citep{Kroupa+1993, Kenyon+1995} and the extinction law \citep{Nishiyama+2009}. The following extinction model was assumed from \cite{Bennett+2015} for dependence on the lens distance, 
\begin{eqnarray}
A_\mathrm{L} = \frac{1-e^{-D_\mathrm{L}\,\mathrm{sin}|b|/0.1\mathrm{kpc}}}{1-e^{-D_\mathrm{RC}\,\mathrm{sin}|b|/0.1\mathrm{kpc}}}A_\mathrm{RC},
\end{eqnarray}
where $A_\mathrm{L}$ is the extinction experienced by the lens and $A_\mathrm{RC}$ is the extinction of the bulge RCGs toward the event coordinate. The equation presumes dust in foreground of the lens with a dust scale height of 0.1 kpc.
In addition to a stellar luminous lens, we also generated a remnant dark lens in accordance with the PARSEC isochrone models \citep{Bressan+2012, Chen+2014, Tang+2014} of stellar evolution and the remnant initial-final mass relation \citep{Lam+2020} implemented in the simulation code. Natal kick velocities of 350 and 100 $\mathrm{km/s}$ are assumed for neutron stars and black holes respectively, following \cite{Lam+2020}.
\par
Figures \ref{fig:5}, \ref{fig:7} and \ref{fig:9} show the posterior probability density distributions of the physical parameters for Model-A, Model-B P$_+$ and Model-B P+LOM$_+$ with $10^6$ simulated microlensing samples accepted under the constraints, and Table \ref{tab:3} lists the physical parameters. We did not simulate events for Model-B P$_-$ and Model-B P+LOM$_-$ since their likelihood and constraints on the Bayesian analysis are practically same as that for Model-B P$_+$ and Model-B P+LOM$_+$, respectively. 
\par
The event simulation results assuming Model-A indicate the lens system is most likely a planetary system consisting of a late-type star orbited by a gas giant. 
The simulation also generated remnant samples; white dwarf lenses at almost identical parameter ranges as main sequence lenses occupy $\sim 70\%$ of the remnant distributions. A bimodal feature appears in the lens distance distribution because the prior probability is weighted more to the Galactic center region whereas the model requires a nearby lens for a low mass system, i.e. main sequence and white dwarf lenses, to explain the longer $t_\mathrm{E}$ and larger $\theta_\mathrm{E}$, which are proportional to the square root of $\pi_\mathrm{rel}=(1\ \mathrm{au})(1/D_\mathrm{L}-1/D_\mathrm{S})$.
\par
On the other hand for Model-B, the result supports a conclusion that the lens system is most likely composed of a remnant orbited by a gas giant, brown dwarf or red dwarf depending on the host mass, which is a reasonable consequence of the longer $t_\mathrm{E}$ and larger $\theta_\mathrm{E}$ likelihood compared to that suggested by Model-A. Especially, it is notable that the predicted percentage of remnant lenses is $100\%$ for Model-B P+LOM$_+$ with more than $95\%$ of the distributions occupied by black hole lenses. Discrepancies between the distributions of the Model-B P$_+$ and Model-B P+LOM$_+$ parameters as well as fractions in the lens types mostly comes from the broad likelihood distribution of the Model-B P$_+$ parameters due to the dispersion in the finite source effect parameter. Upper end of the main sequence distributions are constrained by the upper limit imposed on the lens brightness.


\section{Summary and Discussion} \label{sec:discussion}
The results of the light curve modeling and investigation of the source and lens properties were shown in the preceding sections. We found two degenerate models, Model-A and Model-B, that have similar companion parameter values of $q\sim5\times10^{-3}$ and $s~\sim1.1$ but dissimilar microlensing parameter values of $t_\mathrm{E}\sim(70, 110)$ days and $\rho\sim(5, 1)\times10^{-4}$. Due to the long event timescale and large angular Einstein radius, both models favored a nearby heavy lens solution; furthermore, a Bayesian analysis including remnant populations in the lens system prior revealed the lens to be a remnant candidate.
\subsection{Interpretation of the Lens System}
Although several combinations of the host and companion object types are proposed regarding the posterior distributions as shown in Table \ref{tab:3}, they can be divided roughly into three pairs, a gas giant planet with a main sequence star, a gas giant with a remnant, and a brown dwarf or red dwarf with a black hole (BH).\par
Both close-in and distant giant planets around main sequence stars are found to date by various survey techniques and statistically studied for their occurrence around different spectral types of hosts. 
For microlensing, \cite{Suzuki+2018} compared the statistical analysis result of planet occurrence presented in \cite{Suzuki+2016} to the core accretion model \citep{Ida+2004, Mordasini+2009} and confirmed excess in microlensing planets beyond $q \sim 10^{-4}$, including a factor $\sim 5$ discrepancy for $10^{-3} < q \leq 0.03$. The favored value for $q$ for event OGLE-2014-BLG-0221 will add to that discrepancy. The MOA collaboration is preparing an extended analysis of \cite{Suzuki+2016} with extended data beyond $2007 - 2012$ in which OGLE-2014-BLG-0221 will be included.
There is other observational evidence from various observational methods that support some modification to the traditional planet formation scenarios of giant planets. A recent study of transiting planets discovered by TESS indicates a higher occurrence rate of giant planets around low mass stars than the rate predicted by the core accretion model \citep{Bryant+2023}. The discovery of four distant giant planets around HR 8799 by direct imaging suggests that gravitational instability \citep{Boss+1997} would play an important role in a planet formation \citep{Marois+2008, Marois+2010}. 
\par
There are a few cases of exoplanets discovered around a white dwarf (WD) or a neutron star, including the first exoplanets detected in 1992 around a pulsar, PSR B1257+12 \citep{Wolszczan+1992}. However, most of them represent extreme environment; for example, PSR B1620-26AB b is a circumbinary planet around a pulsar-WD binary \citep{Thorsett+1993, Sigurdsson+2003}, WD 0806-661 b has a very large separation of 2500 au from its host \citep{Luhman+2011} and WD1856+534 b has a small separation of 0.02 au \citep{Vanderburg}. Only MOA-2010-BLG-477Lb is a currently known planet orbiting about a few au away from a WD \citep{Blackman+2021}. It is still uncertain how a solar analogous planetary system evolves along with its host star. Several mechanisms are proposed such as common envelop during the giant phase \citep{Paczynski+1976} that results in short planetary orbits or stellar mass loss that pushes planets outward \citep{Veras+2016}. Second generation exoplanets, those formed during the post main sequence phase, are also proposed as an alternative scenario for some of the observed systems \citep{Peters+2010, Ledda+2023}. Despite its importance of finding more observational samples, observational bias hinders the solid detection of a WD planetary system. The low luminosity of WDs makes astrometry and transit monitoring difficult, the non-characteristic features of a WD spectrum prevents measuring radial velocity, and direct imaging is biased toward wide orbits as WD 0806-661 b shows. In contrast, microlensing has an advantage as it does not rely on host star brightness, and a microlensing survey toward the Galactic center has a peak sensitivity of planet-host separation at a few au, that is suitable for filling the gap between close-in to distant planets around WDs. 
Microlensing can also detect a planet around a X-ray quiet neutron star with its capability of finding dark objects that is not achievable by any other method. Although the pulsar timing method is the only method successful to date for detecting a planet around a neutron star\footnote{7 planets are confirmed including 3 plants belong to the PSR B1257+12 system \citep{Akeson+2013}.}, microlensing would shed light on a cold planet around an unseen neutron star if the host of OGLE-2014-BLG-0221 is a neutron star. \par

More than a dozen stellar mass BH binaries are known but are mostly found by light from X-ray binaries where accretion of material from a companion occurs and is therefore a closely packed system \citep[e.g.][]{Corral-Santana+2016}. A few non-interacting black hole candidates are also identified by radial velocity \citep[e.g.][]{Shenar+2022, Mahy+2022} and astrometry \citep{El-Badry+2023}, in which $\sim 1.4$ au is the widest orbital separation between the candidate and its companion. Several population synthesis studies predict that large fractions of BH luminous companion binaries should have a wide separation such that orbital period becomes more than years \citep[e.g.][]{Chawla+2022}, much longer than any of the confirmed candidates. We estimated OGLE-2014-BLG-0221 has a projected separation of $\sim 10$ au for the case of a black hole host; thus, it would represent the longest BH binary separation ever known and belong to the theoretically predicted population of wide orbit BH binaries. Moreover, the companion is supposed to be a very low mass star or a brown dwarf, and in either case is too faint to be detected by photometry, indicating such a system is unlikely found by any other method besides microlensing.

\subsection{Future Follow-up Observation}
Once the source and lens are separated enough following their proper motions after several years of the event peak, we would be able to resolve the source and lens with high resolution imaging (e.g. Hubble Space Telescope, Keck adaptive optics). This idea was first developed in \cite{Bennett+2006, Bennett+2007}, and the number of identifications has increased in recent years \citep[e.g.][]{Bennett+2015, Bhattacharya+2018, Terry+2021}. Measurement of the lens-source separation and the lens brightness allows strong constraints to be placed on the mass-distance relation of the lens. Here, we consider the detectability of the lens using the future high resolution imaging of OGLE-2014-BLG-0221.\par

\begin{figure}[ht!]
\includegraphics[width=\linewidth]{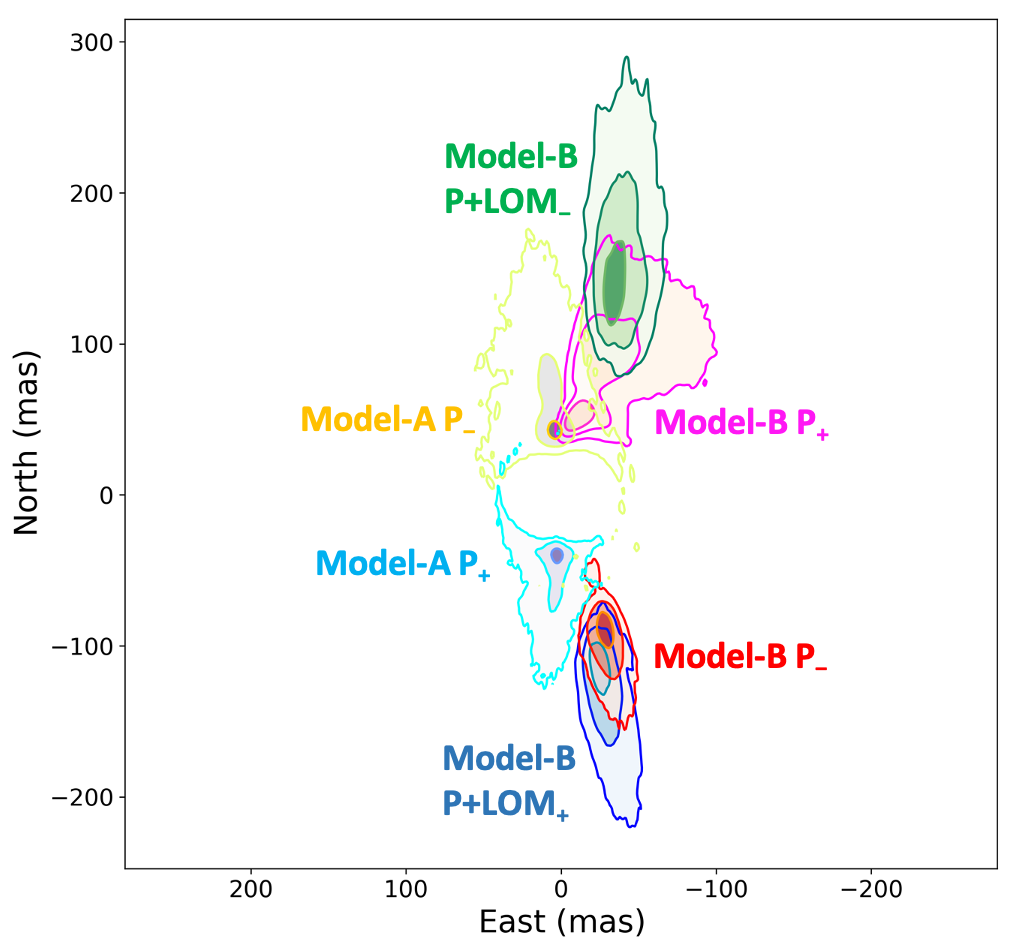}
\caption{Current source-lens relative position. The distributions of Model-A P$_\pm$, Model-B P$_\pm$ and Model-B P+LOM$_\pm$ are plotted. The contour from dark to light shows 39.3, 86.5 and 98.9\% highest density regions.}
\label{fig:10}
\end{figure}

Figure \ref{fig:10} is the posterior distribution of the predicted source-lens separation for each degenerate model with the parallax parameters included, derived from the MCMC parameter chains. 
The separation is computed from the heliocentric proper motion vector:
\begin{eqnarray}
    \bm{\mu_\mathrm{rel, hel}}=\mu_\mathrm{rel}\bm{\hat{\pi}_\mathrm{E}}+\bm{v_\oplus} \frac{\pi_\mathrm{rel}}{\mathrm{au}},
\end{eqnarray}
where $\bm{\hat{\pi}_\mathrm{E}}$ is the unit vector of the microlensing parallax and $\bm{v_\oplus}$ is the Earth's projected velocity at $t_0$. While the direction of the separation is largely dependent on the microlensing parallax vector, the conversion from the geocentric to the heliocentric reference frame does not change the separation itself much unless the relative parallax is too large. We estimated the current separation is $41 \pm 7$ and $45 \pm 14$ mas for Model-A P$_\pm$, $103 \pm 28$ and $98 \pm 13$ mas for Model-B P$_\pm$, and $139 \pm 20$ and $132 \pm 28$ mas for Model-B P+LOM$_\pm$, respectively. 
Hence, we expect the source and lens are separated enough to conduct the high resolution imaging, otherwise the separation is unexpectedly small owing to its uncertainty. By measuring the current position, we would discriminate the models as the direction and magnitude of the separation are likely different among the models. The large expected source-lens separation of Model-B would be due to the small lens distance or the kick velocity caused when a NS or BH forms as discussed in \cite{Lam+2020}.\par
The relative source-lens brightness is also a major concern when resolving the source and lens. Similar brightness is preferred to identify both source and lens; however, \cite{Bhattacharya+2021} demonstrated that the lens can be identified even if it is a few magnitudes fainter than the source. For the case of OGLE-2014-BLG-0221, photometric detection of the lens indicates that the system most likely follows Model-A.
From the Bayesian posterior distribution, we obtained the estimation of the apparent lens magnitude $(H_\mathrm{L}, K_\mathrm{L})$ = $(17.9^{+1.1}_{-1.1}, 17.7^{+1.0}_{-1.1})$ for Model-A. This should be luminous enough as, for example, \cite{Blackman+2021} determined the detection limit of their high resolution Keck image as 21.1 mag in \textit{H}-band. In comparison to the lens brightness, the similar apparent source magnitude is expected from the modeling as $(H_\mathrm{S}, K_\mathrm{S})$ = $(18.1, 17.9) \pm (0.8, 0.8)$ for Model-A, making the event ideal for the high resolution imaging. Furthermore, Model-B measures the different apparent source magnitude, $(H_\mathrm{S}, K_\mathrm{S}) = (18.9, 18.7) \pm (0.8, 0.8)$ and $(18.8, 18.6) \pm (1.2, 1.2)$ for Model-B P$_+$ and Model-B P+LOM$_+$, from that of Model-A. This indicates the model degeneracy would be disentangled with observations of the source even though the lens is not observable for the case of a remnant lens. 
Future high resolution imaging is highly important for characterizing OGLE-2014-BLG-0221.

\begin{acknowledgments}
We would appreciate Iona Kondo and Kento Masuda for valuable comments and discussions.
The MOA project is supported by JSPS KAKENHI Grant Number JP24253004, JP26247023,JP16H06287 and JP22H00153.
RK was supported by JST SPRING, Grant Number JPMJSP2138.
DPB acknowledges support from NASA grants 80NSSC20K0886 and 80NSSC18K0793.
DS was supported by JSPS KAKENHI grant No. 19KK0082.
This work was supported by JSPS Core-to-Core Program (grant number: JPJSCCA20210003). 
NK was supported by the JSPS overseas research fellowship.
\end{acknowledgments}












\bibliography{main}{}
\bibliographystyle{aasjournal}



\end{document}